\documentclass[12pt]{article}
\usepackage{graphicx,color}
\usepackage[utf8]{inputenc}
\usepackage{a4,amssymb,amsmath}

\usepackage{newtxtext}

\numberwithin{equation}{section}

\newtheorem{thm}{Theorem}[section]  
\newtheorem{prop}[thm]{Proposition} 
\newtheorem{lemma}[thm]{Lemma}       
\newtheorem{conj}[thm]{Conjecture}           

\def\eref#1{(\ref{#1})}         
\def\sref#1{Sect.~\ref{#1}}

\def\pref#1{Prop.~\ref{#1}}
\def\lref#1{Lemma~\ref{#1}}

\def\cjref#1{Conj.~\ref{#1}}
\def\cref#1{Cor.~\ref{#1}}

\def\eps{\varepsilon}
\def\ka{\kappa}
\def\la{\lambda}

\def\HH{\mathcal {H}}

\def\pa{\partial}
\def\sumno{\sum\nolimits}
\def\inv{^{-1}}
\def\ol{\overline}
\def\ad{\mathrm{ad}}
\def\lra{\leftrightarrow}

\def\RA{\Rightarrow}

\def\ad{\mathrm{ad}}
\def\Erw#1{\big\langle#1\big\rangle}
\def\erw#1{\langle#1\rangle}
\def\sla{\slash\!\!\!}
\def\wick#1{\,\colon\!#1\!\colon}
\def\wt{\widetilde}

\def\lra{\leftrightarrow}

\def\Sy{\mathfrak S}
\def\eins{\mathbf{1}}
\def\NN{\mathbb{N}}
\def\sig{\sigma}
\def\sfrac#1#2{\hbox{\large{$\frac{#1}{#2}$}}}

\def\bea#1{\begin{eqnarray}\label{#1}}
\def\eea{\end{eqnarray}}
\def\ba{\begin{array}}
\def\ea{\end{array}}
\def\bpm{\begin{pmatrix}} \def\epm{\end{pmatrix}}
\def\ben{\begin{enumerate}}
\def\een{\end{enumerate}}
\def\qbox#1{\quad\hbox{#1}\quad}

\def\vev#1{\langle\!\langle #1\rangle\!\rangle}
\def\erw#1{\langle #1\rangle}
\def\eerw#1{\langle\!\langle #1\rangle\!\rangle}
\def\Erw#1{\big\langle #1\big\rangle}
\def\ERW#1{\Big\langle #1\Big\rangle}
\def\wick#1{\,\colon\!#1\!\colon}

\def\ket#1{\vert #1\rangle}
\def\sla{\slash\!\!\!}

\def\eps{\varepsilon}

\def\sfrac#1#2{\hbox{\large{$\frac{#1}{#2}$}}}
\def\id{i\delta_{xx'}}
\def\Au{\,{}^u\!\!A}
\def\Fu{\,{}^u\!F}
\def\noid{}
\def\Vert{\big\vert}
\def\VERT{\Big\vert}
\def\qed{\hfill $\square$}
\newcommand{\red}[1]{\textcolor{red}{#1}}    
\newcommand{\green}[1]{\textcolor{green}{#1}}  

\title{Non-abelian models in sQFT}
\author{Karl-Henning Rehren$^{1}\quad$%
\footnote{Email: krehren@uni-goettingen.de}
\\[6pt] 
{\footnotesize $^1$Institut für Theoretische Physik, 
Georg-August-Universität Göttingen, 37077 Göttingen, Germany.}
}

\begin{document}

\maketitle

\begin{abstract} It is shown how string-localized QFT predicts interactions of the Standard Model. Detailed computations are presented for various non-abelian models:
  Yang-Mills, QCD, Higgs-Kibble. They serve as backup for \cite{LV,aut} and as
  ``proof of concept'' for the treatment of the full Standard Model. 
  \end{abstract}

\section{Purpose of these notes}
These notes are not (yet) intended for journal publication. They serve as
documentation of model-specific computations backing up claims made in the more conceptual but non-technical paper
\cite{aut} on string-localized QFT (sQFT) and in the paper \cite{LV}
that is concerned with the 
higher-order systematics of perturbation theory in sQFT rather than the actual computations.

Updates of these notes may be found once in a while at
https://www.theorie.physik.uni-goettingen.de/papers/
rehren/24/nab-models.pdf.

String-localized quantum field theory (sQFT) allows to determine
interactions by the constraints imposed by fundamental quantum
principles, and to assess the localization of interacting quantum
fields that in a Hilbert space must be -- by physical necessity -- at variance with the usual
``axiom of locality''. For more background information, see
\cite{aut,LV}. 

The general terminology of sQFT is used without detailed
explanations. General formulas from \cite{LV} are displayed when used.

Much effort is devoted to the appearance of obstructions involving
``string-integrated delta functions'' -- a characteristic feature in
non-abelian models. These obstructions decouple from the determination
of higher-order interactions, but they are essential for the
localization of interacting fields.

\sref{s:QED} (QED) is mainly a streamlined reminder of what has been done in
\cite{Gauss,Infra}, adapted to the needs of \sref{s:QCD} (QCD).
\sref{s:HK} (Higgs-Kibble) is meant as a ``proof of concept'' for the more
involved ongoing analysis of the electroweak interactions \cite{weak}.

\section{Quantum Electrodynamics}
\label{s:QED}

We review sQFT methods developped for QED in \cite{Infra}. Several of
them are also useful in the non-abelian cases. 

\subsection{The autonomous method: $L$-$Q$ pair}
Define $F_{\mu\nu}$ on the Wigner Fock space, and
\bea{Ac} A_\mu(x,c):= \int d\sigma(e)\, c(e) \int_0^\infty ds\,
F_{\mu\nu}(x+se) e^\nu ,
\eea
where $d\sigma(e)$ is the Lorentz invariant measure on the spacelike
hyperboloid $H_1$.
Because $c$ must have unit weight, a variation must be of
the form $\delta c(e)= \pa^e_\ka b^\ka (e)$. It follows
\bea{dcA}\delta_c A_\mu(x,c) &=& \int d\sigma(e)\, b^\ka(e) \int_0^\infty ds\,
\pa^e_\ka (F_{\mu\nu}(x+se) e^\nu) \\ \notag &=& \int d\sigma(e)\, b^\ka(e)
\int_0^\infty ds\, (s\pa^x_\ka F_{\mu\nu}(x+se) e^\nu +
F_{\mu\ka}(x+se)) \\
\notag &=& \int d\sigma(e)\, b^\ka(e)
\int_0^\infty ds\, (s(\pa^x_\mu F_{\ka\nu}(x+se)+\pa^x_\nu
F_{\mu\ka}(x+se)) e^\nu + F_{\mu\ka}(x+se))
\\
\notag &=& \int d\sigma(e)\, b^\ka(e)
\int_0^\infty ds\, \big(s\pa^x_\mu F_{\ka\nu}(x+se)e^\nu+\pa_s(sF_{\mu\ka}(x+se))\big)
.
\eea
The $s$-integral over the second term vanishes, and the first term
gives
\bea{w}\delta_c A_\mu(x,c)  = \pa_\mu w(x,\delta c) 
\quad\hbox{with}\quad w(x,\delta c) = \int d\sigma(e)\, b^\ka(e)
\int_0^\infty s\, ds\, F_{\ka\nu}(x+se)e^\nu.
\eea

The interaction density is
\bea{LcQED}L(c,x) = q\, L_1(c,x)= q\, A_\mu(x,c)j^\mu(x).\eea
The $L$-$Q$ pair is
\bea{LQQED}L_{1,\rm QED}(c) &=& A_\mu(c)j^\mu\\ \notag Q_{1,\rm QED}^\mu&=& w j^\mu.\eea

The second-order obstruction requires to compute
\bea{2po-LQQED} [T,\pa_\mu] j^\mu(x)j^\ka(x')\quad\hbox{and}\quad [T,\pa_\mu] w(x,\delta c),A_\ka(x',c).
\eea
The first vanishes by the standard neutral Ward identity for the Dirac
current. For the second, we use the kinematic propagator (time-ordered
vacuum expectation value) $\vev{TF(x)F(x')}$ and perform the string integrations to define the
kinematic propagators $\vev{Tw X'}$ and $\vev{T\pa w X'}$ for $X=F$ or
$A$. With these
definitions, one has $\vev{T\pa w X'}=\pa\vev{Tw X'}$ because no
field equations are used under the $T$-product which could produce a
delta function when the derivative acts outside, see also \sref{s:QCD}. Consequently, also
the second term in \eref{2po-LQQED} vanishes. It follows that 
\bea{O2-QED}
O^{(2)}(x,x'):=  2\Sy_{xx'} [T,\pa_\mu] Q_1^\mu(x,c,\delta
c)L_1(x',c) = 2\Sy_{xx'} j^\mu(x) \cdot[T,\pa_\mu] w (x,\delta
c)A_\ka(x',c) \cdot j^\ka(x')\eea
vanishes, and the S-matrix is string-independent without induced
interactions.

As discussed in \cite{Infra}, this is true for the S-matrix on the
vacuum Hilbert space, without the extension to superselection sectors
with non-trivial ``photon clouds''. The latter arise due to the
infrared singularity of QED that was not considered here.

\subsection{$L$-$V$ pairs allow comparison with gauge theory}
From now on, we use short-hand
notations for string-integrations. For $e$ spacelike, 
\bea{Inu}
\int_0^\infty ds\,f(x+se) \equiv (I_e f)(x) \quad\hbox{and}\quad \int
d\sigma(e)\, c(e)e^\nu (I_ef)(x) \equiv\Erw{e^\nu
  I_ef}_c(x) \equiv (I_c^\nu f)(x),
\eea
where $c(e)$ has total weight 1. Similarly $I_u^\nu = u^\nu I_u$ and $I_\ell^\nu = \ell^\nu I_\ell$ for
$e=u$ timelike or $e=\ell$ lightlike, in which cases a
smearing (averaging) with $c(e)$ is not necessary. For $f$ vanishing
at infinity it holds
\bea{Ipa}
(\pa_\mu I_v^\mu f)(x) = (I_v^\mu\pa_\mu f)(x) = \int_0^\infty ds\,
v^\nu \pa_\mu f(x+sv) = \int_0^\infty ds\, \sfrac d{ds} f(x+se) = -f(x),\eea
where $v=e,u,\ell$. Because $c(e)$ has total weight 1, the same
identity holds for $I_c$.

To establish that the string-independent S-matrix coincides with that
of gauge theory (in the vacuum sector, and at tree level), we embed
the Maxwell tensor into the indefinite state space (Krein space) of the gauge potential
$A^K_\mu$ in the Feynman gauge, via
\bea{FK} F^K_{\mu\nu} = \pa_\mu A^K_\nu-\pa_\nu A^K_\mu.
\eea
This also embeds $A_\mu(c)$ via \eref{Ac} (with $F$ replaced by
$F^K$) into the Krein space, and it holds 
one has
\bea{Aphi} A^K_\mu(x,c) &=& I^\nu_c(\pa_\mu A^K_\nu-\pa_\nu A^K_\mu) =
A^K_\mu(x) + \pa_\mu\phi(x,c)\quad\hbox{where}\quad \phi(x,c):=(I_c^\nu A^K_\nu). \eea
This yields the $L$-$V$ pair
\bea{LV-K-QED}L^K_{1,\rm QED}(c) &=& A_{\mu}(c)j^\mu, \\ \notag
K_{1,\rm QED}&=& A^K_{\mu}j^\mu, \\ \notag V_{1,\rm
  QED}^\mu(c) &=& \phi(c)j^\mu.\eea
We define kinematic propagators involving $\phi(c)$ and $A^K_\mu(c)$ by
performing the string integrations over the 
Feynman gauge propagator $\vev{TA^K A'^K}$, see \sref{s:QCD}. Again, this makes all
obstructions of interest vanish. There are no induced interactions,
and the S-matrices computed with $L(c)=q\, A(c)j$ and with $q\, A^K j$
coincide.

However, there is an issue with Gauss' Law. The free Maxwell field
$F^K$ satisfies
\bea{dFN} \pa^\mu F^K_{\mu\nu} = -\pa_\nu N \quad\hbox{with}\quad N=
\pa^\mu A^K_\mu,
\eea
known as the ``fictitious current''. $N$ is a null field: its
two-point function vanishes, and so do all 
correlations with $F^K$. But its correlations with $A^K$ do not
vanish. This produces a non-vanishing source term of ``fictitious
provenience'' in charged states created by the Dirac field when $F^{\mu\nu}$
and $\psi$ are perturbed with the local gauge interaction $q\, A^Kj$. The
fictitious source term is cancelled in first order when 
the interaction $q\,\pa_\mu V_1$ is added, see
\cite[Table 1]{Infra}. 

Indeed, the perturbation theory of the fields $F^K$ and $\psi$ with
interaction $L^K(c)$ is the same as with interaction that of  $F$ and $\psi$ with
interaction $L(c)$, because the gauge potential $A^K$ does not
appear in the expansion, and therefore all correlations and
propagators involving the null
field $N$ vanish. By the ``magic formula'' \cite{Infra,aut}, one also gets the same
perturbation theory when one subjects $F^K$ (which is its own dressed
field because all bosonic obstructions vanish) and the dressed field
$\psi_{[q]}$ to the local gauge interaction $q\, A^Kj$. Schematically,
\bea{chain1}\{\psi, F\}\vert_{q\, A(c)j} = \{\psi, F^K\}\vert_{q\, A^K(c)j} = \{\psi_{[q]}, F\}\vert_{q\, A^Kj}.\eea 
The first version is defined on the Wigner Fock space of the electron,
positron and photon, while the locality of the interacting fields
cannot be assessed. The second version is identical with the first, but with
its fields embedded into the Krein space. The third version is defined on the
Krein space, and the perturbatively defined dressed Dirac field is string-localized.
Consequently, also the interacting Dirac field in the first two versions is
string-localized.

\subsection{The dressed Dirac field (formal construction)} The ``magic formula''
\cite{Infra,aut}
\bea{magic} \psi\vert_{q\,A(c)j} = \psi_{[q]}\vert_{q\,A^Kj}
\eea
characterizes properties of the interacting Dirac
field in terms of the ``dressed Dirac field'' as an
intermediate ``free infrafield''.
The perturbative expansion of the latter is \cite{LV}
\bea{dress-pQED} \psi_{[q]}(x) = \exp\big[ iq \int dy\,
O_{V_1(y)}\big](\psi(x)),\eea
where 
\bea{} O_{V_1(y)}(\psi(x)) := \phi(y)\cdot [T,\pa_\mu] j^\mu(y)\psi(x) = \phi(y)\psi(x)\cdot \delta(y-x)
\eea
by the standard charged Ward identity for the Dirac current, see also \sref{s:QCD}. Iterating
$O_{V_1}$ and resumming the exponential series, this results in the
formal expression 
\bea{dress-QED} \psi_{[q]}(x) = \wick{e^{iq\phi(x,c)}}\cdot \psi(x).\eea
However, the field $\phi(c)$ is infrared divergent, so that the
all its Wick powers and the exponential series are ill-defined. In contrast, the exponential
understood as a limit of a Weyl operator can be regularized in the IR, which
provides a non-perturbative definition of \eref{dress-QED} on an
extended Hilbert space with a rich superselection structure
\cite[Sect.~3]{Infra}. The construction requires the support of
admissible smearing functions $c(e)$ to be orthogonal to an arbitrary timelike unit
vector $u$, which ensures that two-point functions among $\phi(c)$ and
$\phi(c')$ define a positive definite inner product.

The magic formula requires to define $\psi_{[q]}$ and the interaction
$q\, A^Kj$ as operators on the same space. This requires an artifice
because $\psi_{[q]}$ is defined on the superselected photon cloud
extension of the Wigner Fock space, while $q\, A^Kj$ is defined on the
Feynman gauge Krein space, and the former cannot be embedded into the
latter. While there is a formal extension
of the superselected Hilbert space to an indefinite space that also
supports $A^K$, the control of positivity remains unclear.

\subsection{The dressed Dirac field (Hilbert space construction)}
This motivates yet another version to define the perturbation theory.
One redefines the Feynman gauge potential by adding a term involving
the null field $N=(\pa A^K)$, so
as to remove the fictitious current for the corresponding Maxwell
field:
\bea{Au} \Au_\mu(x):= A^K_\mu(x) + I_u^\mu (\pa A^K)(x) \quad\hbox{and}\quad \Fu_{\mu\nu} := \pa_\mu \Au_\nu-\pa_\nu \Au_\mu, \eea
where $u$ is an arbitrary timelike unit vector. Then
it holds $\pa \Au=0$ and $\pa \Fu=0$. Moreover, because $\vev{FN}=0$, all correlations
of $\Fu$ are the same as those of $F$.

The field $\Au$ creates from the vacuum only states for which
$(uA^K)^-\ket\Phi=0$. This follows from
\bea{}\vev{A^K_\mu(x) \Au_\nu(x')} =-(\delta_\nu^\ka +
u_\nu I'_u\pa'^\ka) \eta_{\mu\ka}W(x-x') = -(\eta_{\mu\nu} +
u_\nu I'_u\pa'_\mu) W(x-x'),
\eea
and $I'_u(u\pa')=-1$. Such states $\ket\Phi$ generate a positive-definite Fock subspace
with three spacelike massless modes (e.g., $a_i^+(k)\ket0$ if
$u=(1,\vec 0)$ with CCR $[a_i(k),a_j^+(k')] = \delta_{ij}\cdot(2\pi)^32k^0\delta(k-k')$).

Choosing the support of $c(e)$ orthogonal to $u$ and defining
$I_c^\nu$ as in \eref{Inu} with the $SO(3)$-invariant measure on the
sphere $H_1\cap u^\perp$,
one defines the associated
string-localized potential
\bea{Auc} \Au_\mu(x,c):= I_c^\nu(\Fu_{\mu\nu})(x)  = \Au_\mu(x) + \pa_\mu \phi(x,c),\eea
where $\phi(c)$ the same as in
\eref{Aphi} because $e_\mu I_u^\mu=0$.
Then one has another $L$-$V$ pair
\bea{LV-u-QED}L^u_{1,\rm QED}(c) &=& \Au_{\mu}(c)j^\mu, \\ \notag
K^u_{1,\rm QED}&=& \Au_{\mu}j^\mu, \\ \notag V_{1,\rm
  QED}^\mu(c) &=& \phi(c)j^\mu,\eea
and one may write instead of
\eref{chain1}
\bea{chain2}\{\psi, F\}\vert_{q\, A(c)j} = \{\psi, \Fu\}\vert_{q\, \Au(c)j} =
\{\psi_{[q]}, \Fu\}\vert_{q\, \Au j}.\eea
In particular, because $\phi$ does not depend on $u$, the dressed field
constructed with the $L$-$V$ pair \eref{LV-u-QED} is the same as with
the pair \eref{LV-K-QED}.

The advantage of the equivalences \eref{chain2} is that all three
versions are manifestly positive. The disadvantage is that, unlike
$\Fu$, $\Au$ is a non-local field (because 
$I_u$ is a kind of inverse square root of the Laplacian $\Delta_u$ in
the plane orthogonal to $u$, namely on the massless mass-shell it
holds $(u\pa)^2-\Delta_u=\square=0$, hence 
$\Delta_u I_u^2 =(u\pa)^2 I_u^2 = 1$). Then the interaction $q\, \Au j$
is non-local but positive, while the interaction $q\, A^Kj$ in \eref{chain1}
was local but indefinite. The equivalence of the resulting
perturbative expansions has been tested in a nontrivial instance \cite[Sect.~4]{Infra},
where all non-local terms involving $I_u$ cancel out identically.

\section{Yang-Mills and Quantum Chromodynamics}
\label{s:QCD}

We want to compute the induced interactions for YM and QCD (= YM with
massless quarks), and in particular confirm that there are no induced
interactions at third order. This can be done with an $L$-$Q$ pair.

We also want to compute the dressed quark field, for which we also
need an $L$-$V$ pair on a Hilbert space. The latter requires the
artifice previously used in QED.

$L$-$Q$ pairs involving several massless fields of helicity 1 (``free gluons''),
described by Maxwell tensors $F^a_{\mu\nu}$ ($a=1,\dots,N$) as well as
their string-localized potentials $A^a_{\mu}(c)$,
necessarily involve coefficients $f_{abc}$ which are the completely
anti-symmetric structure constants of some Lie algebra \cite{YM}, see
below.

We assume a simple non-abelian Lie
algebra with generators satisfying
\bea{Lie} [\xi_a,\xi_b]= if_{abc} \xi_c,\eea
and we define the symmetric inner product of the Lie algebra
\bea{cart} \Erw{\xi_a\Vert \xi_b} = \delta_{ab}.\eea
It enjoys the invariance property $\erw{gXg\inv\vert
  gZg\inv}=\erw{X\vert Y}$, implying
\bea{cart-inv} \Erw{i[X,Z]\Vert Y} = \Erw{X\Vert i[Z,Y]},
\quad\hbox{in particular}\quad \Erw{i[X,Y]\Vert Y}=0. 
\eea
We shall write interactions in terms of Lie-algebra valued
fields $F_{\mu\nu} = F_{\mu\nu}^a\xi_a$ and $A_\mu(c) = A_\mu^a(c)\xi_a$, and extend the
notations \eref{Lie} and \eref{cart} to Lie-algebra valued
fields, where Wick ordering of the operators is always understood.
E.g., $i[F,A] = -f_{abc} \wick{F^aA^b} \xi_c$.

\subsection{$L$-$Q$ and $L$-$V$ pairs}
\label{s:LQV-QCD}

The $L$-$Q$ pair for YM is 
\bea{LQ-YM}L_{1,\rm YM}(c)&=& \sfrac12\Erw{F^{\mu\nu}\Vert
  i[A_\mu(c),A_\nu(c)]}\\ \notag Q_{1,\rm YM}^\mu&=&
\Erw{F^{\mu\nu}\Vert i[w,A_\nu]}\equiv-\Erw{i[F^{\mu\nu},A_\nu(c)]\Vert
  w}.
\eea
The cubic Yang-Mills interaction in the Feynman
gauge is
\bea{K1-YM} K_{1,\rm YM}=\sfrac12\Erw{F^{K,\mu\nu}\Vert i[A^K_\mu,A^K_\nu]}.
\eea

However, embedding $F$ in the Krein space as in \eref{FK}, there is no
$L$-$V$ relating the resulting $L^K_{1,\rm YM}(c)$ to the cubic
gauge theory interaction \eref{K1-YM} in the Feynman gauge, because
the difference is not a derivative (suppressing
irrelevant common factors and the superscript $K$ for the sake of this computation):
\bea{L-K?}L_1(c)-K_1&\sim& \Erw{F^{\mu\nu}\Vert
  [A_\mu(c)-A_\mu,A_\nu(c)+ A_\nu]} =\Erw{F^{\mu\nu}\Vert
  [\pa_\mu\phi(c),A_\nu(c)+ A_\nu]}  \\ \notag &=& \pa_\mu\Erw{F^{\mu\nu}\Vert
  [\phi(c),A_\nu(c)+ A_\nu]} - \Erw{\pa_\mu F^{\mu\nu}\Vert
  [\phi(c),A_\nu(c)+ A_\nu]} - \Erw{F^{\mu\nu}\Vert
  [\phi(c),\pa_\mu A_\nu(c)+ \pa_\mu A_\nu]} .
\eea
The last term equals $- \erw{F^{\mu\nu}\vert
  [\phi(c),F_{\mu\nu}}=0$ by \eref{cart-inv}, but the second term
exhibits the
``fictitious current'' \eref{dFN}:
\bea{fict} \Erw{\pa^\nu N\Vert [\phi,A_\nu(c)+A_\nu]}=\pa^\nu\Erw{N\Vert [\phi,A_\nu(c)+A_\nu]} - \Erw{N\Vert
  [\pa^\nu \phi,A_\nu(c)+A_\nu]} - \Erw{N\Vert [\phi,\pa^\nu A_\nu(c)+\pa^\nu A_\nu]},\quad
\eea
where the last term equals $-\erw{N\vert [\phi,N]}=0$ by \eref{fict} and
\eref{cart-inv}, and the second term equals
\bea{}-2\Erw{N\Vert [\pa^\nu\phi,A_\nu]}=-2\pa^\nu\Erw{N\Vert
  [\phi,A_\nu]}+ 2\Erw{\pa^\nu N\Vert [\phi,A_\nu]} + 2\Erw{N\Vert [\phi,N]}.
\eea
Again, the last term vanishes but the second term does not. Thus,
$L_1(c)-K_1$ is not a derivative.

We shall therefore proceed with vector potentials $\Au^{a}$
($a=1,\dots,N$) as introduced in QED, so that $\pa \Fu^{a}=0$. Then,
redoing \eref{L-K?} with $\Au$, we have the $L$-$V$ pair for YM
\bea{LV-YM}
L_{1,\rm YM} &=& \sfrac12\Erw{\Fu^{\mu\nu}\Vert
  i[\Au_\mu(c),\Au_\nu(c)]},\\ \notag
K_{1,\rm YM} &=& \sfrac12\Erw{\Fu^{\mu\nu}\Vert
  i[\Au_\mu,\Au_\nu]},\\ \notag
V^\mu_{1,\rm YM} &=& \Erw{\Fu^{\mu\nu}\Vert
  i[\phi,\Au_\nu(c)+\Au_\nu]}.
\eea

The $L$-$Q$ and $L$-$V$ pairs for QCD are obtained by adding
\bea{LQV-QCDmin}
L_{1,\rm QCD,min}=\Erw{\Au_\mu(c)\Vert j^\mu},
\\ \notag
Q^\mu_{1,\rm QCD,min}=\Erw{w\Vert j^\mu}, \\ \notag
K_{1,\rm QCD,min}=\Erw{\Au_\mu\Vert j^\mu},\\ \notag
V^\mu_{1,\rm QCD,min}=\Erw{\phi(c)\Vert j^\mu}
\eea
to \eref{LQ-YM} and \eref{LV-YM}, where $j^\mu_a = \ol \psi \gamma^\mu
\pi(\xi_a)\psi$.

\subsection{Obstructions} All obstructions of the S-matrix can be
computed in terms of ``obstruction maps''
\bea{OYX}
O_{Y(x)}(X(x')) := T[\pa^x_\mu Y^\mu(x) X(x')]-\pa^x_\mu T[Y^\mu(x)
X(x')],  \quad \hbox{in short}\quad O_Y(X')= [T,\pa_\mu] Y^\mu
X',
\eea
where $X$ and $Y^\mu$ are Wick polynomials. For our purposes, tree
level is sufficient, and is always understood. Then by Wick's theorem,
\bea{OYXW} O_Y(X') =\sum_{\varphi,\chi} \sfrac{\pa
  Y^\mu}{\pa\varphi}(x) O_\mu (\varphi(x),\chi(x'))\sfrac{\pa
  X}{\pa\chi}(x') \equiv \sum_{\varphi,\chi} \sfrac{\pa
  Y^\mu}{\pa\varphi}O_\mu (\varphi,\chi')\sfrac{\pa
  X'}{\pa\chi'}  \eea
where the sum runs over all linear free fields appearing in $Y$ and
$X$, and the ``two-point obstructions'' are defined in terms of the
the propagators = time-ordered two-point functions
\bea{}O_\mu (\varphi(x),\chi(x'))  = \vev{T[\pa^x_\mu \varphi(x) \chi(x')]}-\pa^x_\mu \vev{T[\varphi(x)
\chi(x')]}\equiv
\vev{[T,\pa_\mu] \varphi\chi'}.
\eea
They may be non-vanishing whenever in the first term some equation of
motion is used under the $T$-product, or when non-kinematic
propagators are used, as will be necessary in \sref{s:HK}.

\subsection{Propagators and two-point obstructions} All obstruction maps
are computed in terms of the propagators of free fields.
We use the shorthand notation $T_m\equiv T_m(x-x')$ for the time-ordered
two-point function of a scalar field of mass $m$ (the Feynman
propagator is $-iT_m$), and $\delta_{xx'}\equiv \delta(x-x')$.

We begin with the free Dirac field of mass $m$ ($m=0$ for QCD, and
$m=m_e$ in the Higgs-Kibble model in \sref{s:HK}, which we shall also 
put to $m_e=0$).
$$\vev{\gamma^\mu[T,\pa_\mu]\psi\ol\psi{}'} = -im \vev{T\psi\ol\psi{}'} -
\sla\pa\vev{T\psi\ol\psi{}'} =
-i(m-i\sla\pa)(m+i\sla\pa)T_m=-i(\square+m^2)T_m = -\delta_{xx'},$$
and similarly
$$\vev{[T,\pa_\mu]\psi'\ol\psi\gamma^\mu} = -\gamma^0 \big(\gamma^\mu[T,\pa_\mu]\psi\ol\psi{}'\big)^*\gamma^0 = \delta_{xx'}.$$
For several independent Dirac fields $\psi_m$ in a representation
$\pi$ of a
Lie algebra, we have to multiply with
$\delta_{mn}$.

We now turn to (non-abelian) currents
$j_a^\mu = \ol \psi \gamma^\mu \pi(\tau_a)\psi$ with Lie algebra generators
$\xi_a$. (The latter are $\tau_0=1$ for QED, $\tau_a=\frac12 \lambda_a$ for QCD, and
$\tau_a=\frac12\sigma_a$ for $\mathfrak{su}(2)$.)
$$[T,\pa_\mu]j_a^\mu\psi'_m
=([T,\pa_\mu]\psi'_m\ol\psi_n\gamma^\mu)\pi(\tau_a)_{nk}\psi_k
=\delta_{mn}\delta_{xx'} \pi(\tau_a)_{nk}\psi_k = (\pi(\tau_a)\psi)_m\cdot\delta_{xx'},
$$
and similarly
$$[T,\pa_\mu]j_a^\mu\ol\psi{}'_m
=-(\ol\psi\pi(\tau_a))_m\cdot\delta_{xx'}.
$$
Consequently,
$$[T,\pa_\mu]j^\mu_aj'^\ka_b =
-(\ol\psi\pi(\tau_a))\gamma^\ka\pi(\tau_b)\psi'\cdot\delta_{xx'} +
\ol\psi{}'\gamma^\ka\pi(\tau_b)(\pi(\tau_a)\psi)\cdot\delta_{xx'} =
-\ol\psi\pi([\tau_a,\tau_b])\gamma^\ka\psi'\cdot\delta_{xx'} = -f_{abc}
j^\ka_c\cdot i\delta_{xx'}.
$$
In short:
\bea{Ojj} O_\mu(j^\mu_a,j'^\ka)= -f_{abc}j^\ka_c\id.
  \eea

\paragraph{Two-point obstructions for $L$-$Q$.}
We proceed with massless gluon fields. All fields carry a Lie-algebra
index $a=1,\dots,N$, and all their correlations have a
factor $\delta_{ab}$  (not displayed).

Starting from the Wigner field $F$
\bea{TFF}
\vev{TF_{\mu\nu}(x)F_{\ka\la}(x')} =
-\pa_{[\mu}\eta_{\nu][\ka}\pa_{\la]} T_0(x-x'), 
\eea
we compute the kinematic propagators involving $A_\mu(c)$ (given by
\eref{dcA}), needed in $L$-$Q$:
\bea{LQprop}
\vev{TF_{\mu\nu}A'^\ka(c)} &=& -(\pa_{[\mu}\delta_{\nu]}^\ka - \pa_{[\mu}I'_{c,\nu]}\pa^\ka)T_0,\\ \notag
\vev{TA_\mu(c)F'^{\ka\la}} &=& -(\delta_\mu^{[\ka}\pa^{\la]} +\pa_\mu
I_c^{[\ka}\pa^{\la]})T_0 ,\\ \notag
\vev{TA_\mu(c)A'^\ka(c)} &=& -( \delta_\mu^\ka +\pa_\mu
I_c^\ka - I'_{c,\mu} \pa^\ka -(I_cI_c')\pa_\mu \pa^\ka )T_0,
\eea
where $I'^\nu_c$ is the string integration w.r.t.\ the argument $x'$.
We may renormalize \eref{TFF} by adding $-c_F
\eta_{\mu[\ka}\eta_{\la]\nu}\cdot \id$. Then we might impose that
time-ordering commutes with $I_c$, which (by definition of $A(c)$)
produces also renormalizations of \eref{LQprop}. As shown in
\cite{GassPhD}, this option would produce within $O^{(2)}$ a non-resolvable obstruction
due to $O_\mu(A_\nu,A'^\ka)$, involving $F F'$ and a string-localized
delta function. Its absence forces $c_F=0$. On the other hand, in
view of \eref{Ipa}, time-ordering should rather not commute with $I_c$
because it does not commute with $\pa$. Then, the renormalizations of
\eref{LQprop} are independent, and can be chosen zero, while still
keeping the renormalization of \eref{TFF}. This amounts to the
prescription (adopted later) that only propagators of local fields
should admit renormalizations, which can only be (derivatives of)
delta functions, constrained by the scaling dimension. We shall return
to this issue later, when the third order is dicussed.

Kinematic propagators involving $w$ (given by \eref{w}) and $\pa w = \delta_cA_\mu(c)$
can be computed as well, but it is sufficient to know that one gets
$\vev{T\pa wX'}=\pa\vev{TwX'}$.

Using the equations of motion
\bea{eomLQ} \pa^\mu F_{\mu\nu}=0,\quad \pa_\mu A_\nu(c) - \pa_\nu A_\mu(c) = F_{\mu\nu},\quad \pa^\mu A_\mu(c)=0,
\eea
under the $T$-product, one computes the relevant
two-point obstructions 
\bea{Table1}\begin{tabular}{l||c|c|c|c|}
Gluon & $F'^{\ka\la}$ &$A'^{\ka}(c)$  \cr\hline\hline
$O_\mu(F^{\mu\nu},\cdot)$
& $ -(1+c_F)\eta^{\nu[\ka}\pa^{\la]}\id$&$-(\eta^{\nu\ka}-I'^\nu_c\pa^\ka)\id$\cr
     $O_{[\mu}(A_{\nu]}(c),\cdot)$  &$-c_F\delta_\mu^{[\ka}\delta_\nu^{\la]}\id$&$0$ \cr 
$O_\mu(A^\mu(c),\cdot)$  &$-I_c^{[\ka}\pa^{\la]}\id$&$-\big(I_c^\ka -(I_cI_c')\pa^\ka\big)\id$ \cr
$O_\mu(w,\cdot)$ &$0$&$0$\cr  \hline
\end{tabular}\quad
\eea
{\bf Table 1.} Two-point obstructions in the gluon sector
(Lie-algebra indices suppressed, all entries times $\delta_{ab}$)

\paragraph{Two-point obstruction for $L$-$V$.}
Starting from the Krein space field $A^K$
\bea{}
\vev{TA^K_{\mu}(x)A^K_{\ka}(x')} =
-\eta_{\mu\ka} T_0(x-x'), 
\eea
we compute the kinematic propagators involving $\Fu$ and $\Au$
(identical to \eref{LQprop}, because $N$ is a null field and the Cauchy-Schwarz
inequality holds on the Hilbert space) as well as
$\Au(c)$ and $\phi(c)$, needed in $L$-$V$:
\bea{LVprop}
\vev{T\Au_\mu \Au{}'{}^\ka} =-(\delta_\mu^\ka + I_{u,\mu} \pa^\ka - \pa_\mu
I'^\ka_u)T_0, &\quad& \vev{T\phi(c)\phi'(c)} = -(I_cI_c') T_0, \\ \notag
\vev{T\Au_\mu \phi'(c)} = -(I'_{c,\mu} + I_{u,\mu})T_0, &\quad& \vev{T\phi(c)\Au'^\ka} = -(I_c^\ka  +
I'^\ka_u)T_0, \\ \notag
\vev{T\Fu_{\mu\nu}\Au'^\ka} = -(\pa_{[\mu}\delta_{\nu]}^\ka + \pa_{[\mu}I_{u,\nu]} \pa^\ka )T_0, &\quad& \vev{T\Au_\mu \Fu'^{\ka\la}}
=-(\delta_\mu^{[\ka}\pa^{\la]} -\pa_\mu
I'^{[\ka}_u\pa^{\la]})T_0, \\ \notag
\vev{T\Fu_{\mu\nu}\phi'(c)} = -(\pa_{[\mu}I'_{c,\nu]} +
\pa_{[\mu}I_{u,\nu]})T_0, &\quad& \vev{T\phi(c) \Fu'^{\ka\la}} = -(I_c^{[\ka}\pa^{\la]} +I'^{[\ka}_u\pa^{\la]})T_0, 
\\ \notag
\vev{T\Au_\mu (c)\Au'^{\ka}} = -( \delta_\mu^\ka  + \pa_\mu
I_c^\ka + I_{u,\mu}\pa^\ka)T_0,  &\quad& \vev{T\Au_\mu \Au'^{\ka}(c)} = -(\delta_\mu^{\ka}-I'_{c,\mu}\pa^{\ka} -\pa_\mu
I'^{\ka}_u)T_0, \\ \notag
\vev{T\Au_\mu (c)\phi'(c)} = -( I'_{c,\mu} + I_{u,\mu} + (I_cI'_c)\pa_\mu  )T_0,&\quad& \vev{T\phi(c)\Au'^{\ka}(c)} = -(I_c^{\ka}+
I'^{\ka}_u-(I_cI'_c)\pa^{\ka} )T_0.\eea
Using the equations of motion \eref{eomLQ} (with $A(c)$ and $F$
replaced by $\Au(c)$ and $\Fu$) and in addition
\bea{eomLV} \pa_\mu \Au_\nu - \pa_\nu \Au_\mu = \Fu_{\mu\nu},\quad \pa^\mu \Au_\mu=0, \quad
\pa_\mu \phi(c) = \Au_\mu(c)-\Au_\mu,
\eea
under the $T$-product, one computes the relevant
two-point obstructions 
\bea{Table2}\begin{tabular}{l||c|c|c|c|}
Gluon & $\Fu'^{\ka\la}$ &$\Au'^{\ka}(c)$ &$\Au'^{\ka}$ &$\phi'$  \cr\hline\hline
$O_\mu(\Fu^{\mu\nu},\cdot)$
& $ -(1+c_F)\eta^{\nu[\ka}\pa^{\la]}\noid$&$-(\eta^{\nu\ka}-I'^\nu_c\pa^\ka)\noid$&$-(\eta^{\nu\ka}+I^\nu_u\pa^\ka)\noid$& $-(I'^\nu_c+I_u^\nu) \noid$\cr  $O_{[\mu}(\Au_{\nu]}(c),\cdot)$  &$-c_F\delta_\mu^{[\ka}\delta_\nu^{\la]}\noid$&$0$&$0$&$0$ \cr    
 $O_{[\mu}(\Au_{\nu]},\cdot)$  &$-c_F\delta_\mu^{[\ka}\delta_\nu^{\la]}\noid$&$0$&$0$&$0$ \cr  
$O_\mu(\Au^\mu(c),\cdot)$  &$-I_c^{[\ka}\pa^{\la]}\noid$&$-\big(I_c^\ka -(I_cI_c')\pa^\ka\big)\noid$&$-I_c^\ka \noid$&$-(I_cI'_c)\noid$ \cr$O_{\mu}(\Au^\mu,\cdot)$ &$I'^{[\ka}_u\pa^{\la]}\noid$&$I'^\ka_u\noid$&$I'^\ka_u\noid$&$0$  \cr
$O_\mu(\phi,\cdot)$ &$0$&$0$&$0$&$0$ \cr  \hline
\end{tabular}\quad
\eea
{\bf Table 2.} Two-point obstructions in the gluon sector
(Lie-algebra indices suppressed, all entries are displayed as
operators to be  applied to $\id$.)

\subsection{$L$-$Q$ at second order}
Throughout this section, we shall write $A(c)\equiv A$ and $L_1\equiv
L_1$ and $I\equiv I_c$ etc.\ because there
is no ambiguity.

We have to compute and
resolve the second-order obstruction \cite{LV} 
\bea{O2} O^{(2)}(x,x')&=& 2\Sy_{xx'} O_{Q_1}(L_1')\equiv2\Sy_{xx'}\sum_{\varphi,\chi}\ERW{\sfrac{\pa Q_1^\mu}{\pa \varphi}(x)\VERT
  O_\mu(\varphi(x),\chi(x'))\sfrac{\pa L_1}{\pa\chi}(x')}\notag\\ &\stackrel{!}=&\delta_cL_2(x,c)\cdot \id-i\Sy_{xx'}\pa^x_\mu Q_2^\mu(x;x').
\eea
We split $O^{(2)}=O^{(2)}\vert_\delta+O^{(2)}\vert_{I\delta}$ according
to the contributions with delta functions and with 
string-integrated delta functions in \eref{Ojj} and Table 1. The resolution \eref{O2}
requires that $O^{(2)}\vert_{I\delta}$ must arrange into a total derivative. That this is
indeed the case, is a structural feature of the field content of YM
and QCD, which we conveniently formulate in two lemmas.

\begin{lemma}\label{l:L1} If $\delta_cL_1(c)=\pa Q_1$ where
  $L_1$ is a   Wick polynomial in the fields $A_a$ and string-independent
  fields, and $Q_1$ is linear in $u_a$, then
\bea{L1} \hbox{\rm (i)}\quad \frac{\pa L_1}{\pa A_{\mu}} = \frac{\pa Q^\mu_1}{\pa w}, \qquad
\hbox{\rm (ii)}\quad \pa_\mu \Big(\frac{\pa
  Q^\mu_1}{\pa w}\Big)=0.
\eea
\end{lemma}

{\em Proof:} The comparison of
$$\delta_cL_1=\Erw{\frac{\pa L_1}{\pa A^\mu}\Vert\pa^\mu u}$$
with
$$\pa_\mu Q_1^\mu = \pa_\mu \Erw{\frac{\pa Q_1^\mu}{\pa u}\Vert u}
=\Erw{\pa_\mu \Big(\frac{\pa Q_1^\mu}{\pa u}\Big)\Vert u}+\Erw{\frac{\pa
  Q_1^\mu}{\pa u}\Vert \pa_\mu u} $$
immediately yields (i) and (ii). \qed

\begin{lemma}\label{l:ww2}  Both for YM and for QCD, $O^{(2)}\vert_{I\delta}$ is a total derivative. The
corresponding field $Q_2^\mu\vert_{I\delta}(x,x')$ arises from $Q^\mu_1(x)$ by a simple
replacement of $w(x)$ by $2w_2(x,x')$:
\bea{lem2} Q_2^\mu\vert_{I\delta}(x,x') = 2 \Erw{\frac{\pa Q^\mu_1}{\pa
    w}(x)\Vert w_2(x,x')}, \qquad\hbox{where}\quad w_2(x,x'):=i[w',A'_\nu ]\cdot I^\nu\delta_{xx'}.
\eea
\end{lemma}
{\em Proof:} 
By inspection of $Q_1^\mu$ and Table 1, string-integrated delta
functions arise only from $O(F,A' )$. Thus,
$$O_{Q_1}(L'_1)\vert_{I\delta}=\Erw{i[w,A_\nu ]\Vert
  \frac{\pa L'_1}{\pa A'^\ka }}\cdot I'^\nu\pa^\ka\id=-i\pa'_\mu\big(\Erw{i[w,A_\nu ]\Vert
  \frac{\pa Q'^\mu_1}{\pa w'}}\cdot I'^\nu\delta_{xx'}\big),$$
where we have used $\pa\id=-\pa'\id$ and 
\lref{l:L1}. When $(x\lra x')$ is added, \eref{O2} implies the
claim. \qed

We now turn to Yang-Mills, given by \eref{LQ-YM}.
With the apriori result of \lref{l:ww2}, we may concentrate on $O^{(2)}\vert_\delta$. Table 1 gives
\bea{OQ1L1YM} O_{Q_1(L_1')}\vert_\delta&=&
-\Erw{i[w,A_\nu ] \Vert
   i[A'_{\ka} ,A'_{\la} ]}\cdot (1+c_F)\eta^{\nu\ka}\pa^{\la}\id \\ \notag &&+\Erw{i[w,A_\nu ] \Vert
  i[F'_{\ka\la},A'^\la ]}\cdot \eta^{\nu\ka}\id+
\sfrac12\Erw{ i[w,F^{\mu\nu}]\Vert
   i[A'_\ka ,A'_\la ]
}\cdot c_F\delta_\mu^\ka\delta_\nu^\la\id. \eea
The second and third term combine, with the help of the Jacobi
identity, to
\bea{23}\sfrac12(1+c_F) \Erw{i[w,F^{\ka\la}] \Vert
 i[A_{\ka} ,A_{\la} ]}\cdot\id .\eea
When we integrate the first term by parts, we get a term
that exactly cancels \eref{23}, and the remaining terms
$$O_{Q_1(L_1')}\vert_\delta=(1+c_F)\pa^\ka\big(\Erw{i[w,A_\la ] \Vert
 i[A'_{\ka} ,A'_{\la} ]}\id\big)-(1+c_F)\Erw{i[\pa_\ka w,A_\la ] \Vert
 i[A_{\ka} ,A_{\la} ]}.$$
Symmetrizing and adding $O^{(2)}\vert_{I\delta}$ from \lref{l:ww2}, we see:

\begin{prop}\label{p:LQ2-YM} For YM, \eref{O2} is solved
with
\bea{LQ2-YM}L_{2,\rm YM} &=& -\sfrac12(1+c_F)\Erw{i[A^\mu ,A^\nu ]\Vert
  i[A_\mu ,A_\nu ]},\\ \notag Q_{2,\rm YM}^\mu&=& -2(1+c_F)\Erw{i[A^\mu ,A^\nu ]\Vert
  i[w,A_\nu ]}\cdot \delta_{xx'} -2 \Erw{i[F^{\mu\nu},A_\nu ]\Vert w_2(x,x')}.
\eea
\end{prop}
Observe that it holds
\bea{Q2L2} Q_2^\mu\vert_{\delta}(x,x') = \Erw{\frac{\pa L_2}{\pa
    A_\mu }\Vert w}\cdot \delta_{xx'}.
\eea

Turning to QCD, we have to add \eref{LQV-QCDmin}. This adds $O_{Q_{1,\rm
    min}}(L'_{1,\rm min})$ and the ``cross-terms''
$O_{Q_{1,\rm YM}}(L'_{1,\rm min})+O_{Q_{1,\rm min}}(L'_{1,\rm YM})$ to \eref{OQ1L1YM}.
The former is easily computed with \eref{Ojj}:
$$O_{Q_{1,\rm
    min}}(L'_{1,\rm min}) = \Erw{i[w,A_\nu ]\Vert j^\nu}\cdot \id.$$
It is not resolvable. $O_{Q_{1,\rm YM}}(L'_{1,\rm min})$ is zero by
Table 1. $O_{Q_{1,\rm min}}(L'_{1,\rm YM})$ is computed with Table 1:
$$O_{Q_{1,\rm min}}(L'_{1,\rm YM}) =- \Erw{i[w,A_\nu ]\Vert
  j'^\ka}\cdot(\delta^\nu_\ka - I'^\nu\pa_\ka)\id. $$
The $\delta^\nu_\ka$-term cancels the previous (``lock-key''), and the
$I'^\nu\pa_\ka$-term is a total derivative because $\pa=-\pa'$ and $j'$ is
conserved. Thus,

\begin{prop}\label{p:LQ2-QCD} For QCD, \eref{O2} is solved by adding to
\eref{LQ2-YM} the terms
\bea{LQ2-QCD-min}L_{2,\rm QCD, min}&=&0, \\ \notag 
Q_{2,\rm QCD, min}^\mu(x;x')&=& 2\Erw{j^\mu(x)\Vert
  w_2(x,x')},
\eea
with $w_2$ as given in \lref{l:ww2}.
\end{prop}

\subsection{$L$-$Q$ at third order}
We continue writing $A(c) \equiv A$ etc.\ because there
is no ambiguity.

We have to compute and resolve the
third-order obstruction \cite{LV}
\bea{O3}O^{(3)}(x,x',x'') = 3\Sy_{xx'x''}\big(O_{Q_1}(L_2')\cdot
\delta_{x'x''}+O_{Q_2(x;x')}(L_1'')\big) \stackrel{!}=\delta_cL_3\cdot \delta_{xx'x''}-\Sy_{xx'x''}\pa_\mu Q_3^\mu(x;x',x'').\eea
We only want to show that it is a total derivative, so that $L_3=0$.

Let us split
\bea{}O_{Q_2(x;x')}(L_1'')\vert_{I\delta}=\Erw{\sfrac{\pa Q_2}{\pa w_2}
  \Vert O(w_2,L_1'')} +  + \Erw{w_2\Vert O(\sfrac{\pa Q_2}{\pa w_2} ,L_1'')}.
\eea
Then, in \eref{O3}, we can isolate the delta-terms, and, using
\lref{l:ww2} to express $Q_2\vert_{I\delta}$,  arrange the
string-delta terms as follows:
\bea{} O_{Q_1}(L_2')\cdot
\delta_{x'x''}+O_{Q_2(x;x')}(L_1'') &=& (O_{Q_1}(L_2')\vert_{\delta}+O_{Q_1}(L_2')\vert_{I\delta})\cdot
\delta_{x'x''}+O_{Q_2(x;x')\vert_\delta}(L_1'') +
O_{Q_2(x;x')\vert_{I\delta}}(L_1'') \notag \\ &=&
\label{O3loc}
\big[O_{Q_1}(L_2')\vert_\delta\cdot \delta_{x'x''}
+O_{Q_2(x;x')\vert_\delta}(L_1'')\vert_\delta\big] \\
\label{O3w2}
&& + \Sy_{xx'x''}\big[ \Erw{w_2\Vert O(\sfrac{\pa Q_2}{\pa w_2} ,L_1'')} +
O_{Q_1}(L_2')\vert_{I\delta}\cdot\delta_{x'x''}\big] \\
\label{O3remain}
&& + \Sy_{xx'x''} \big[\Erw{\sfrac{\pa Q_2}{\pa w_2}
  \Vert O(w_2,L_1'')}+O_{Q_2(x;x')\vert_{\delta}}(L_1'')\vert_{I\delta}\big].
\eea
It is clear that \eref{O3loc} determines $L_3-\pa
Q_3\vert_\delta$.

\begin{prop}\label{p:L3-QCD}
  \eref{O3loc} is identically zero. Thus, provided \eref{O3w2} and \eref{O3remain}
  are derivatives, the third-order obstruction is resolved with
  $L_3=0$ and $Q_3\vert_\delta=0$. 
  \end{prop}
{\em Proof:} The first term in \eref{O3loc} is readily computed with
Table 1, giving a multiple of
$$\Erw{i[w,A_\mu]\Vert i[A_\nu,i[A^\mu,A^\nu]]} =
\Erw{i[i[w,A_\mu],A_\nu]\Vert i[A^\mu,A^\nu]}=\sfrac12 \Erw{i[w,i[A_\mu,A_\nu]]\Vert i[A^\mu,A^\nu]}=0.$$
by \eref{cart-inv} and Jacobi and again \eref{cart-inv}. In the second
term, $Q_2\vert_\delta$ contains a factor $\delta_{xx'}$. This factor
cannot be taken out of
the time-ordered product. Instead, we can take advantage of the symmetrizations:

\begin{lemma}\label{l:delta} It holds
$$\Sy_{xx'}O_{Y\cdot \delta_{xx'}}(X'') = \sfrac12 O_{Y}(X'')\cdot \delta_{xx'}.$$
\end{lemma}
{\em Proof:}
$$2\Sy_{xx'}O_{Y\cdot \delta_{xx'}}(X'') =T (\pa_x+\pa_x')(Y\delta_{xx'})X''- (\pa_x+\pa_x')TY\delta_{xx'}X''.$$
Now, we use $(\pa_x+\pa_x')\delta_{xx'}=0$, hence
$(\pa_x+\pa_x')(f(x)\delta_{xx'})=\delta_{xx'}\cdot(\pa_x+\pa_x')f(x)
=\delta_{xx'}\cdot\pa f(x)$, and get
$$2\Sy_{xx'}O_{Y\cdot \delta_{xx'}}(X'') =T (\pa Y)\cdot\delta_{xx'}X''- (\pa_x+\pa_x')(\delta_{xx'}TYX'') =
\delta_{xx'}\cdot (T\pa Y X'' - \pa TYX'').$$
This proves the Lemma. \qed

{\em Proof of \pref{p:L3-QCD} (continued):} Thanks to \lref{l:delta}, in order to prove the vanishing of the second term in
\eref{O3loc}, it suffices to consider
$O_{Y}(L_1')\vert_\delta$ with $Y^\mu=\erw{w\vert
  i[A_\nu,i[A^\mu,A^\nu]]}$. This requires to know two-point
obstructions $O_\mu(A_\nu,X')\vert_\delta$ for $X'=A',F'$,  of which only the
anti-symmetric and trace parts are already specified in Table
1. But $O_\mu(A_\nu,A'^\ka)$ has scaling dimension 3, so it can
contain only string-integrated delta functions.
$O_\mu(A_\nu,F'^{\ka\la})$ has scaling dimension 4, hence
$O_\mu(A_\nu,F'^{\ka\la})\vert_\delta$ must be a multiple of
$\delta_{\mu}^{[\ka}\delta_\nu^{\la]}\id$, i.e., it coincides with the
anti-symmetric part $\frac12 O_{[\mu}(A_{\nu]},F'^{\ka\la})$ fixed by Table 1.
The computation of $O_{Y}(L_1')\vert_\delta$ is again a multiple of
$\Erw{i[w,i[A_\mu,A_\nu]]\Vert i[A^\mu,A^\nu]}=0$. \qed

For the first of the other two terms, we have an apriori cancellation:
\begin{lemma}\label{l:aprio} \eref{O3w2} is a total derivative.
\end{lemma}
{\em Proof:} We freely permute $x,x',x''$ and suppress the
symmetrization symbols $\Sy$. By \lref{l:ww2}, the first term in \eref{O3w2} equals 
\bea{aprio0}
2 \Erw{w_2\Vert O(\sfrac{\pa Q_1}{\pa w} ,L_1'')} = 2\Erw{w_2(x,x')\Vert
  \sfrac{\pa O_{Q_1}(L_1'')}{\pa w(x)}}.\eea 
Here we use the second-order resolution \eref{O2} together with \eref{Q2L2}:
  $$2\Erw{w\Vert
  \sfrac{\pa O_{Q_1}(L_1')}{\pa w}} =  \Erw{\pa
    w\Vert \sfrac{\pa L_2}{\pa A }}\id
  - \pa\big(\Erw{w\Vert\sfrac{\pa L_2}{\pa A }}\id+Q_2\vert_{I\delta}\big) = -\Erw{w(x)\Vert \pa\sfrac{\pa L_2}{\pa A }}\id 
  - \pa' Q_2(x';x)\vert_{I\delta}.$$
  In the last term, we have swapped $x\lra x'$ so that, by \lref{l:ww2}, it contains
  only $w(x)$ only through
  $w_2(x',x)=i[w(x),A(x)]\cdot\dots$. The derivative does not act on
  $w(x)$. Thus, \eref{aprio0} gives
\bea{aprio1}2\Erw{w_2(x,x')\Vert
  \sfrac{\pa O_{Q_1}(L_1'')}{\pa w(x)}}&=& \Erw{w_2(x,x')\Vert-
  \pa_\mu\sfrac{\pa L_2}{\pa A_\mu }(x)}\cdot i\delta_{xx''}
  - \pa''_\mu\Erw{w_2(x,x')\Vert \sfrac{\pa Q^\mu_2(x'';x)}{\pa w(x)}}).\eea
  On the other hand, the second term in \eref{O3w2} is computed with Table 1:
\bea{aprio2} 
\Erw{i[w',A'_\nu ]\Vert \sfrac{\pa
    L_2}{\pa A_\mu }\cdot I^\nu\pa'_\mu\id}\cdot\delta_{xx''}
=-\Erw{\pa_\mu w_2(x,x')\Vert \sfrac{\pa
    L_2}{\pa A_\mu }}\cdot i\delta_{xx''}.
\eea
Taken together, and symmetrized by $\Sy_{xx'x''}$, the two terms yield
a total derivative.
\qed

It is apparent from the proof that the discarded contribution to $Q_3(x,x',x'')$ will involve terms with a factor
$w_3(x,x',x'')= i[w_2(x',x''),A_\nu(x')] \cdot I^\nu\delta_{xx'}=
i[i[w(x''),A_\mu(x'')] ,A_\nu(x')] \cdot
I^\nu\delta_{xx'}I^\mu\delta_{x'x''}$. However, we shall not care for
$Q_3$ because we only want to establish that $L_3=0$.

With \pref{p:L3-QCD} and \lref{l:aprio}, it remains to establish that
\eref{O3remain} is a derivative, in order for the third-order obstruction to be resolvable.
If this is true, then $L_3=0$ by \pref{p:L3-QCD}. In particular, there is no restriction on the renormalization constant
$c_F$, which appears in \eref{LQ2-YM} as a factor $1+c_F$ multiplying the
quartic YM interaction. This situation is
reminiscent of scalar QED, where there is also a renormalization
constant $c$, and the induced quartic interaction comes with a coefficient
$1+c$ where gauge theory would expect the coefficient $1$.

(In the PGI approach of Scharf et al, it was proven at all perturbative
orders including loops, that the
coefficient of scalar QED can be absorbed in a renormalization group
transformation. A similar result was proven in the sQFT approach to
scalar QED at all orders at tree level (\cite{Tipp}). In the present
case, we don't have an all-orders result yet.)

It remains to establish that \eref{O3remain} is a derivative. 
Here, the problems are more severe than in \eref{O3loc}:
namely, \lref{l:delta} does not deal with $O_{Y(x')\cdot
  I\delta_{xx'}}(X'')$ because $I\delta_{xx'}$ is not symmetric. And
the unspecified two-point obstructions 
$O_\mu(A_\nu,X')\vert_{I\delta}$ cannot be deduced with the arguments
as in the proof of \pref{p:L3-QCD}.

The subsequent ``interlude'' (\sref{s:interlude}) will show that there are at least two options
to ensure that \eref{O3remain} is a total derivative, both somewhat
artificial.  We shall formulate a conjecture that
neither of these artifices is in fact necessary.

\subsection{Interlude: Lessons from $L$-$Q$ and strategies for $L$-$V$}
\label{s:interlude}
We continue writing $A(c) \equiv A$ etc.\ as long as there
is no ambiguity.

Thanks to \lref{l:aprio} and \pref{p:L3-QCD}, it remains to make sure
that \eref{O3remain} is a total derivative:
\bea{remain}
\Sy_{xx'x''} \big[\Erw{\sfrac{\pa Q_2}{\pa w_2}
  \Vert
  O(w_2,L_1'')}+O_{Q_2(x;x')\vert_{\delta}}(L_1'')\vert_{I\delta}\big]
\stackrel!{\stackrel{\rm der}=} 0.
\eea

\paragraph{Option (A).} We first notice that $w_2(x,x') =
i[w',A'_\nu ]\cdot I^\nu\delta_{xx'}$ would vanish if the smearing
with $c(e)$ were sharp, because $I_e^\nu = e^\nu I_e$ and ``axiality''
holds for
\bea{ax-e} e^\nu
A_\nu(e)=0\eea
by definition of $A_\nu(e) = I_e^\nu F_{\mu\nu}$.
That is, in
the sharp case, the problematic first term in \eref{remain} would be
absent.

(Also with smeared strings a version of ``axiality'' holds:
\bea{ax-c} I^\nu
A_\nu =0\eea
by definition of $A_\nu  = I^\nu F_{\mu\nu}$, but this does not
imply the vanishing of $w_2(x,x')$. Even when $w_2$ is integrated over
$x'$, one gets $I^\nu(i[w,A_\nu ])$ which does not vanish because
$I^\nu$ does not act on the two factors separately.)

\lref{l:delta} deals with the second term in \eref{remain}, but its evaluation
still requires to know the string-delta parts of
$O_\mu(A_\nu,X')$. \pref{p:axial} motivates the following choice: 
\bea{OAX}\notag
O_\mu(A_\nu ,A'^\ka ) &:=& 
\sfrac32 (I'_{\mu}\pa_\nu+I'_{c,\nu} \pa_\mu- \eta_{\mu\nu}) I^\ka \id +\sfrac14\eta_{\mu\nu} (II')\pa^\ka\id \\  O_\mu(A_\nu ,F'^{\ka\la}) &:=& - \sfrac12
c_F \delta_\mu^{[\ka}\delta_\nu^{\la]} - \sfrac14\eta_{\mu\nu} I^{[\ka}\pa^{\la]} \id ,
\eea
which amounts to the renormalization of the propagators:
\bea{TpaAX}
\vev{T\pa_\mu A_\nu X'} &=& \pa_\mu
\vev{TA_\nu A'^\ka } + O_\mu(A_\nu ,X').
\eea
It is easily checked that \eref{OAX} are consistent with the second and third lines of Table
1, which were computed by using the equations of motion
\eref{eomLQ} under the $T$-product. In particular, \eref{TpaAX} are consistent with \eref{eomLQ}.

(The choice \eref{OAX} is not unique: \pref{p:axial} also holds when one adds to
$O_\mu(A_\nu ,A'^\ka )$ a multiple of the traceless symmetric structure
$(2I_{(\mu}I_{\nu)}-\eta_{\mu\nu}(II'))\pa^\ka\id$.)

\begin{prop}\label{p:axial} With \eref{OAX}, the second term
  $O_\mu(Q_2^\mu\vert_\delta,L_1'')$ in
\eref{remain} is a derivative, up to terms containing a string-integrated delta
function $I^\nu\delta$ contracted with fields $A_\nu $. In particular, if computed with a sharp string, \eref{remain} is a
derivative and \pref{p:L3-QCD} applies.
\end{prop}

{\em Proof:}
Let $Y^\mu=\erw{w\vert   i[A_\nu,i[A^\mu,A^\nu]]}$. By \lref{l:delta} we have to show that $O_{Y}(L'_1)\vert_{I\delta}$ is a derivative up to a
``violation of axiality'', i.e., terms involving contractions of
factors $A_\nu $ and $I^\nu\id$. We have to compute
\bea{OQ2X} O_{Y}(X') =
i[A^\nu,i[w,A_\nu]] \cdot O_\mu(A^\mu,X') -
\big(i[w,i[A^\mu,A^\nu]]+ i[A^\mu,i[w,A^\nu]]\big)\cdot O_\mu(A_\nu,X')
\quad\eea
and contract with $\frac{\pa L_1'}{\pa X'}$.

$X'=A' $: The $(II')$-part of $O_\mu(A_\nu,A'^\ka)$ in
\eref{OAX} contains a factor $\pa^\ka\id=-\pa'^\ka\id$ which turns the
contraction with  $\frac{\pa L'}{\pa
  A'^{\ka}} = j^\ka-i[F'^{\ka\la},A'_\la]$ into a derivative
because the latter is conserved (\lref{l:L1}). The other part of
$O_\mu(A_\nu,A'^\ka)$ contributes to \eref{OQ2X} as
\bea{}
i[A^\nu,i[w,A_\nu]]\cdot (3-4)I^\ka\id - i[A^\mu,i[w,A^\nu]] \cdot
\sfrac32I'_{(\mu}\pa_{\nu)} I^\ka\id + i[A^\mu,i[w,A^\nu]]\cdot
\eta_{\mu\nu}I^\ka\id.
\eea
The first and the last term cancel, and the middle term vanishes up to
a violation of axiality.

$X'=F'$: The $\eta$-part of $O_\mu(A_\nu,F'^{\ka\la})$ in
\eref{OAX} involves factors $I^\ka$
or $I^\la$. Hence, it vanishes up to another violation of axiality when
contracted with $\frac{\pa L'}{\pa
  F'_{\ka\la}} = \frac12i[A'^{\ka},A'^\la]$.
The anti-symmetric part of $O_\mu(A_\nu,F'^{\ka\la})$ contributes to
$O_Y(L_1')\vert_{\delta}$ and was treated already in the proof of
\pref{p:L3-QCD}. This proves the first claim.

In particular, if evaluated with a sharp string,
the second term in \eref{remain} is a derivative, while $w_2=0$ by
axiality, hence the first term is absent. 
This proves the statement. \qed

We have thus resolved $O^{(3)}$ with a sharp string, where both terms in
\eref{remain} vanish separately. Unfortunately,
the smearing of the linear fields is necessary when one proceeds to
loop level, see the discussion in \cite[Chap.~5]{GassPhD}. With smeared strings, the second term vanishes
only up to a ``violation of axiality'', and for the first term, we
lack a prescription, similar to \lref{l:delta},
how the deal with $O_{Y'\cdot I\delta}(X'')$. We therefore make a conjecture:

\begin{conj}\label{cj:optA} The (unknown) correct prescription for
the evaluation of
$O_{w_2}(\cdot)= O_{i[w',A'_\nu]\cdot I^\nu\delta_{xx'}}$ is such that after symmetrization
\eref{remain}, computed with the
obstructions \eref{OAX} for smeared strings, is a total derivative.
\end{conj}
(In other words: the first term cancels the ``violations of axiality''
in the second term which are absent with a sharp string. The symmetrization is presumably necessary because it was also
necessary in \lref{l:delta}.)

\paragraph{Option (B).} We now propose an alternative, more radical strategy, whose motivation
comes from its application in the $L$-$V$ approach, as discussed
below.

\begin{conj}\label{cj:optB} It is legitimate to redefine the propagators
\bea{TpaAX0} \vev{T\pa_\mu A_\nu X'}:= \pa_\mu \vev{TA_\nu X'},\eea
which implies that all two-point obstructions
$O_\mu(A_\nu ,X')=0$.
\end{conj}

The redefinition \eref{TpaAX0} brings drastic simplifications:
the first term in \eref{remain} vanishes without knowing a precise prescription for its
evaluation simply because $w_2$ involves only the ``inert'' fields $w$
and $A $, and the second term vanishes without 
any computation because $Q_2\vert_\delta$ also involves only $w$ and
$A $. Also \pref{p:L3-QCD} is true without any computation. On the other hand, \eref{TpaAX0} violates the equation of motion
$\pa^\mu A_\mu =0$ under the $T$-product. The 
issue of \cjref{cj:optB} is whether this ``anomaly'' can be tolerated.

The $L$-$V$ treatment of string-localized QCD, outlined in the next
paragraph, gives strong indications that in
fact, both \cjref{cj:optA} and \cjref{cj:optB} are true.
Yet, we think that the last word about the issue is not spoken.

\paragraph{Strategies for $L$-$V$.}

Because we are interested in a possible ``kinematic confinement
criterion'' \cite{aut} that can be assessed at the intermediate level of the
``dressed quark field'', we need a treatment of  sQCD in $L$-$V$. This
shall be done in \sref{s:QCD-LV2} and \sref{s:QCD-LV3}. The
$L$-$V$ pair was already given in \sref{s:LQV-QCD}.

The analysis at second order can be done without problems with Table
2 and allows to compute the dressed quark field at second order (joint
work with M. Chantreau \cite{Cha23}). However, at third order, the problem of evaluating obstructions of the form
$O_{Y'\cdot I\delta}(X'')$ re-appears. It could be avoided with a sharp
string, as above. One then obtains a dressed quark
field with a sharp string, generalizing \eref{dress-QED} to the non-abelian
case.  Namely, we can check that until the third order in the
perturbative approach, the exponential $\wick{e^{ig\phi}}$ of
\eref{dress-QED} is replaced by a
path-ordered exponential $\wick{Pe^{ig\phi}}$, where $\phi$ is the integral over the Lie-algebra valued field
$\Au_\mu$ along the string from $x$ to infinity, and the path-ordering orders the Lie-
algebra generators, while the field operators are Wick-ordered.

However, the ``kinematic confinement criterion'' must be a
non-perturbative IR effect, which requires a regularization of the
path-ordered exponential. Since this was possible in the abelian case
(\cite{Infra}, see \sref{s:QED}) only with a smearing, we need an
alternative. We have the two options which in $L$-$Q$ led to the
same result.

Option (A) is to work with a sharp string and ``extrapolate'' the
final results to smeared strings. With a sharp string, $w_2$ is
identically zero, and it can be left out before evaluating obstructions.
The main justification in $L$-$V$ is this. Notice that the path-ordered exponential with a
sharp string is a Wilson operator, so that the dressed quark field is
invariant when classical gauge transformations are applied to  the
quantum fields.

In the spirit of Dirac's introduction of gauge-invariant
Dirac fields dressed with smeared exponents which are precisely the classical
versions of the dressed Dirac fields \eref{dress-QED} of sQED, one
should then ``extrapolate'' the non-abelian case to smeared strings by
demanding that the dressed quark field also remains gauge-invariant
(under gauge transformations that are trivial at infinity). This
``pseudo-classical'' approach contradicts the ``autonomous'' attitude,
but it predicts the higher-order terms in the exponential by a
recursive system of equations. Its iterative solution coincides 
at second order with the term computed with $L$-$V$, and predicts a string-localized third-order term, that in the case of a sharp string
also coincides with the $L$-$V$ term, \sref{s:dressQCD}.

Option (B) is to proceed in analogy with \eref{TpaAX0} and re-define
\bea{OuAX0}
O_\mu(\Au_\nu(c),X'):=0 \qbox{and} O_\mu(\Au_\nu,X'):=0
\eea
(which does not change the second-order results but violates
the equations of motion $\pa^\mu \Au_\mu(c)=\pa^\mu \Au_\mu=0$ under the $T$-product), and take advantage
of the resulting drastic simplifications. It makes $w_2$ ``inert'',
i.e., $O(w_2,X')=0$, and cancellations occur in the evaluated obstructions. 

In \sref{s:QCD-LV3}, we shall use Option (B). The computations are still
quite intricate because we do not know the $L$-$V$ analogue of the apriori
cancellations as in \lref{l:aprio}; but the fact that very substantial cancellation will occur
at in paragraph 3.1 and 3.2 in \sref{s:QCD-LV3} suggests that such an analogue should exist.

Option (B) is justified in retrospective by some remarkable facts: while terms involving
string-integrations $I_u$ abound during the computation, there is a
complete cancellation of them in the third-order
result for the dressed quark field; and moreover this result coincides
with the ``extrapolation of the pseudo-classical expression'', mentioned
in the Option (A).

Given these findings at third-order, there still remains a lot to
do: (i) find a closed expression for the ``smeared Wilson operator''
in all orders (e.g., by developing a notion of (classical) ``smeared parallel
transport'' in GT language and establishing the expression with the
$L$-$V$ method), (ii) find a non-perturbative definition of the
quantum smeared Wilson operator generalizing
the abelian one, and (iii) test confinement criteria in terms of
correlation functions of quantum smeared Wilson operators, see \sref{s:kinc}

\subsection{$L$-$V$ at second order.}
\label{s:QCD-LV2}
Throughout this section, we shall drop the superscript ${}^u$ and
write $\Au(c)\equiv A(c)$, $\Au\equiv A$ and $L^u_1\equiv
L_1$, $K^u_1\equiv K_1$ etc.\ because there is no ambiguity. For
simplicity, we put $c_F=0$.

The $L$-$V$ methods recursively determines higher-order interactions
$L_n(x,c)$ and $K_n(x)$ along with fields $U_n^\mu(x;x_1\cdots
x_{n-1})$ \cite{LV}, where $U_1\equiv V_1$.
We shall show that the former vanish for $n\geq 3$. The latter will be
needed to compute the ``dressed quark field'' at order $n$. For the time being, we cannot go beyond $n=3$. 

The subsequent computations were completed before we became aware of
\lref{l:ww2} in $L$-$Q$ to compute the string-delta part at second
order, and of the
systematic apriori cancellations at third order, \lref{l:aprio}. There
are certainly also $L$-$V$ versions of these 
results, which would make not least the lengthy paragraphs 3.1 and 3.2
in \sref{s:QCD-LV3} below obsolete.

To determine the second-order interactions, we have to compute and
resolve the second-order obstruction \cite{LV} 
\bea{P2}P^{(2)}(x,x')=\Sy_{xx'} O_{U_1}(L_1'+K_1')& \stackrel!=& (L_2(x,c)-K_2(x))\id - i\Sy_{xx'}\pa
_\mu U_2^\mu(x;x').\eea
The $L$-$V$ pair was specified in \sref{s:LQV-QCD}, and the two-point
obstructions are specified in Table 2. 

We shall drop the superscript ${}^u$ throughout, and use some convenient abbreviations: the conserved currents 
\bea{JJ} J_\ka(c):= \frac{\pa L_1}{\pa
  A^\ka(c)} =j_\ka-i[F_{\ka\la},A^\la(c)]\qquad\hbox{and}\quad J_\ka:=\frac{\pa K_1}{\pa
  A^\ka} =j_\ka-i[F_{\ka\la},A^\la],
\eea
\bea{AA} A_{\ka\la}(c) := 2\frac{\pa L_1}{\pa
  F^{\ka\la}} = i[A_\ka(c),A_\la(c)] \qquad\hbox{and}\quad A_{\ka\la} := 2\frac{\pa K_1}{\pa
  F^{\ka\la}}= i[A_\ka,A_\la],
\eea
and 
\bea{sig} \sig_\nu := 2\frac{\pa V_1^\mu}{\pa
  F^{\mu\nu}} \equiv i\big[\phi,
\frac{\pa(L_1+K_1)}{\pa j^\nu}\big]= i[\phi,A_\nu(c)+A_\nu]  .\eea

$O_{U_1}$ involves only the anti-symmetric parts of the two-point obstructions 
$O_{\mu}(A_{\nu}(c),\cdot)$ and $O_{\mu}(A_{\nu},\cdot)$. These vanish
in Table 2. Thus, one needs only the contributions from $\varphi=F$
and $\varphi=j$, giving
\bea{Fj}
O_{U_1}(L'_1+K'_1)=\ERW{\sfrac12\sig_\nu\VERT
O_\mu(F^{\mu\nu},\chi')\frac{\pa(L'_1+K'_1)}{\pa\chi'}} +
\Erw{\sig_\ka\Vert j^\ka}\id.\eea
Specifically, the $F$-contribution to $O_{U_1}(L_1'+K_1')$ is
$$\Erw{\sfrac12\sig_\nu \Vert
  (-\eta^{\nu[\ka}\pa^{\la]})\id \cdot \sfrac
  12(A'_{\ka\la}(c)+A'_{\ka\la})-(\eta^{\nu\ka}-I_c'^\nu\pa^\ka)\id
  \cdot J'_\ka(c)-(\eta^{\nu\ka}+I_u'^\nu\pa^\ka)\id \cdot J'_\ka}$$
Rearranging the manifest derivative terms of the $F$-contribution
$$\pa^\la\Erw {\sfrac 12\sig_\nu\Vert -2\eta^{\nu\ka}\id
  \cdot \sfrac
  12(A_{\ka\la}(c)+A_{\ka\la})} +\pa'^\ka\Erw{\sfrac
  12\sig_\ka\Vert -I_c'^\nu\id \cdot J'_\ka(c) +
  I_u^\nu\id\cdot J'_\ka},$$
produces the first three terms of $U_2^\mu$, given in
\eref{LKU2-QCD}. The remaining terms  in \eref{Fj}
are purely gluonic because 
the $j$-parts of the $F$-contribution cancel the $j$-contribution
$\erw{\sig\vert j}\id$. They can be written as $i\delta_{xx'}$ times
$$-\sfrac12 \Erw{\pa^\kappa i[\phi(c),A^\la(c)+A^\la]\Vert A_{\ka\la}(c)+A_{\ka\la}}+
\sfrac 12\Erw{\sig_\ka\Vert i[F^{\ka\la},A_\la(c)+A_\la]}.$$
In the first term, letting $\pa^\ka$ act on
$\phi(c)$, where $\pa \phi = A(c)-A$, we get
\bea{}
-\sfrac12\Erw{i[A^\ka(c)-A^\ka, A^\la(c)+A^\la]\Vert A_{\ka\la}(c)+A_{\ka\la}}
&=& -\sfrac12\Erw{A_{\ka\la}(c)-A_{\ka\la}\Vert
  A_{\ka\la}(c)+A_{\ka\la}} \notag \\
&=& -\sfrac12\Erw{A_{\ka\la}(c)\Vert A^{\ka\la}(c)}+ \sfrac12\Erw{A_{\ka\la}\Vert A^{\ka\la}}.\eea
These terms give rise to $L_2-K_2$ in \eref{LKU2-QCD}.
\begin{prop}\label{p:L2-QCD}
All remaining terms in \eref{Fj} are total derivatives, giving rise to the
last term in $U_2$ in \eref{LKU2-QCD}.
\end{prop}
{\em Proof:} The remaining terms are $\id$ times
$$-\sfrac12 \Erw{i[\phi(c),\pa^\kappa A^\la(c)+\pa^\kappa A^\la]\Vert
  A_{\ka\la}(c)+A_{\ka\la}}+
\sfrac 12\Erw{i[\phi(c),A_\ka(c)+A_\ka]\Vert i[F^{\ka\la}, A_\la(c)+
A_\la]} = $$$$=-\sfrac12\Erw{\phi(c)\Vert 
[F^{\ka\la},A_{\ka\la}(c)+A_{\ka\la}]  - i[A_\ka(c)+A_\ka,i[F^{\ka\la}, A_\la(c)+
A_\la]]}=:\sfrac12\Erw{\phi(c)\Vert X}.$$
Here,
$$X=[F^{\ka\la},[A_\ka(c),A_\la(c)]+[A_\ka,A_\la]] - \sfrac12
[A_\ka(c)+A_\ka,[F^{\ka\la}, A_\la(c) + A_\la]] +\sfrac12 [A_\la(c)+A_\la,[F^{\ka\la}, A_\ka(c) + A_\ka]]\big)$$
$$  = [F^{\ka\la},[A_\ka(c),A_\la(c)]+[A_\ka,A_\la]] - \sfrac12[F^{\ka\la},[A_\ka(c)+A_\ka,A_\la(c) + A_\la]]$$
$$  = \sfrac12[F^{\ka\la},[A_\ka(c)-A_\ka,A_\la(c) - A_\la]]  \equiv  \sfrac12[F^{\ka\la},[\pa_\ka\phi,\pa_\la\phi]].$$
On the other hand,
$$\pa_\ka[\phi,[\phi,\pa_\la\phi]]-(\ka\leftrightarrow\la) =\underbrace{[\pa_\ka\phi,[\phi,\pa_\la\phi]]-[\pa_\la\phi,[\phi,\pa_\ka\phi]]}_{=[\phi,[\phi_\ka,\phi_\la]]}
+ 2 [\phi,[\pa_\ka\phi,\pa_\la\phi]] = 3[\phi,[\pa_\ka\phi,\pa_\la\phi]] .$$
Therefore,
$$\Erw{\phi\Vert X}=\sfrac12
\Erw{[\phi,F^{\ka\la}]\Vert[\pa_\ka\phi,\pa_\la\phi]}=-\sfrac12
\Erw{[F^{\ka\la},\phi]\Vert [\pa_\ka\phi,\pa_\la\phi]}=-\sfrac12
\Erw{F^{\ka\la}\Vert[\phi,[\pa_\ka\phi,\pa_\la\phi]]}=$$
$$=-\sfrac16
\Erw{F^{\ka\la}\Vert\pa_\ka[\phi,[\phi,\pa_\la\phi]]-(\ka\leftrightarrow\la)}
= -\sfrac13\pa_\ka
\Erw{F^{\ka\la}\Vert[\phi,[\phi,\pa_\la\phi]]} =:\pa_\ka Y^\ka.$$ 
Then
$$\Erw{\phi\Vert X}\cdot\delta_{xx'} = (\pa_\mu Y^\mu)\cdot\delta_{xx'} =
((\pa_\mu+\pa'_\mu) Y^\mu)\cdot\delta_{xx'} =
(\pa_\mu+\pa'_\mu) (Y^\mu\cdot\delta_{xx'})=\Sy_{xx'}\pa_\mu(2Y^\mu\cdot\delta_{xx'})$$
is a derivative. The corresponding part of $U_2$ is
$-Y^\mu\cdot\delta_{xx'}$. \qed

This yields
\bea{LKU2-QCD}
L_{2,\rm QCD} &=& -\sfrac12 \Erw{A^{\mu\nu}(c) \Vert A_{\mu\nu}(c)},\notag\\
K_{2,\rm QCD} &=& -\sfrac12 \Erw{A^{\mu\nu}\Vert A_{\mu\nu}},\notag\\
U^\mu_{2,\rm QCD} &=& \sfrac12 \Erw{J^\mu(c)\Vert \sig'_\ka}\cdot
I_c^\ka\delta_{xx'} -\sfrac12 \Erw{J^\mu\Vert \sig'_\ka}\cdot
I_u'^\ka\delta_{xx'} \notag \\
&&  
-\sfrac12
\Erw{A^{\mu\nu}(c)+A^{\mu\nu}\Vert \sig_\nu}\cdot\delta_{xx'}
-\sfrac13
\Erw{F^{\mu\nu} \Vert i[\phi,i[\phi,\pa_\nu\phi]]}\cdot\delta_{xx'}. 
\eea

\subsection{$L$-$V$ at third order.}
\label{s:QCD-LV3}
We continue dropping the superscript ${}^u$ because there is no ambiguity.

\paragraph{1. Preparations.} We repeat the analysis of
string-independence in $L$-$V$ -- firstly in order to show equivalence
with a local interaction from gauge theory, and secondly, in order to
be able to compute and analyze the ``dressed quark field'', whose
IR properties are expected to have a bearing on confinement, see \sref{s:kinc}.

We expect that $L_n=K_n=0$ for $n>2$. We are going to check this for
$n=3$ (paragraph 3 and 4). For that purpose, it suffices to show that 
$P^{(3)}$ is a total derivative. On the other hand, for the
dressed field (\sref{s:dressQCD}), we need the quark part of the derivative terms
$\pa U_3$ (paragraph 5). In particular, total derivatives of purely gluonic terms may
be ignored for these purposes.

One has to compute and resolve (see \cite{LV})
\bea{P3}P^{(3)}(x,x',x'')&=& \Sy_{xx'x''}\big(\sfrac12 O_{U_1}O_{U'_1}(L''_1-K''_1) -\sfrac{3i}2
\big(O_{U_2}(L''_1+K''_1)+O_{U_1}((L'_2+K'_2)\delta_{x'x''})\big) \\
&\stackrel!=&(L_3-K_3)\delta_{xx'x''} -\Sy_{xx'x''} \pa^x U_3(x;x',x''). \notag\eea
We choose the arguments $x$, $x'$,
$x''$ conveniently, and at some points freely permute them, because we
have to symmetrize anyway. 

The $I\delta$-parts of $U_2$ cause in $O_{U_2}(L_1''+K_1'')$ the same
kind of trouble as $w_2$ in $Q_2\vert_{I\delta}$ in $L$-$Q$. As
explained in \sref{s:interlude}, we avoid the trouble, by adopting Option (B):
we set all obstructions $O_\mu(A_\nu(c), X')=0$ and $O_\mu(A_\nu,
X')=0$, without changing the second-order results.
Thus, 
we shall have, generalizing \eref{Fj}
\bea{Fjgen} O_Y(X') = \sum_{\chi'} \ERW{\frac{\pa Y^\mu}{\pa
    F^{\mu\nu}}\VERT \frac{\pa X'}{\pa \chi'}} \cdot
O_\mu(F^{\mu\nu},\chi') + \ERW{i\Big[ \frac{\pa Y^\mu}{\pa
    j^{\mu}}, \frac{\pa X'}{\pa
    j'^\ka}\Big]\VERT j^\ka}\cdot \id.
\eea

For the evaluation of some contributions to \eref{P3}, we will have to
apply \lref{l:delta}. For the iteration of obstruction maps
$O_{U_1}\circ O_{U'_1}(\cdot)$ it we be 
convenient to use another lemma.

\begin{lemma}\label{l:Opa} It holds
\bea{Opa} O_{U'}(\pa Y) = \pa O_{U'}(Y) -\pa'O_{Y}(U')+
O_{Y}(\pa'U').\eea
\end{lemma}

{\em Proof:}
\eref{Opa} is equivalent to
$[[T,\pa'],\pa]U(x')Y(x)=[[T,\pa],\pa']U(x')Y(x)$, which follows from $[\pa,\pa']=0$. \qed

\paragraph{2. Notations.}
It is convenient to introduce some more notation:
$$S_\ka := A_\ka(c)+A_\ka=\pa_\ka\phi, \quad D_\ka = A_\ka(c)-A_\ka,$$
$$S_{\ka\la}:=A_{\ka\la}(c)+A_{\ka\la}
=\sfrac12\big(i[D_\ka,D_\la]+i[S_\ka,S_\la]\big),\quad D_{\ka\la}:=A_{\ka\la}(c)-A_{\ka\la} =
\sfrac12\big(i[D_\ka,S_\la]+i[S_\ka,D_\la]\big)
,$$
\bea{pasig}\sig_\nu = i[\phi,S_\nu] \quad\RA\quad \pa^{[\ka}\sigma^{\la]}=
2(D^{\ka\la}+ i[\phi,F^{\ka\la}]).\eea

We also write
\bea{Ipm} I_{xx'}^{\pm\nu} := (I_c\pm I'_u)^\nu\delta_{xx'}.
\eea
Then the relevant quantities are
\bea{LKU1} L_1+K_1 &=& \sfrac12\Erw{F^{\ka\la}\Vert S_{\ka\la}} +
\Erw{S_\ka\Vert j^\ka}, \notag\\
L_1-K_1 &=& \sfrac12\Erw{F^{\ka\la}\Vert D_{\ka\la}} +
\Erw{D_\ka\Vert j^\ka}, \notag\\
U_1^\mu &=& \sfrac12\Erw{F^{\mu\nu}\Vert \sig_\nu} +
\Erw{\phi \Vert j^\ka},
\eea
\bea{LKU2} L_2+K_2  &=& -\sfrac14
\Erw{S^{\ka\la}\Vert S_{\ka\la}} -\sfrac14 \Erw{D^{\ka\la}\Vert D_{\ka\la}},
\notag\\
U_2^\mu &=& \sfrac12 \Erw{\sig'_\nu \Vert j^\mu}\cdot I^{-\nu}_{xx'}
-\sfrac14\Erw{i[F^{\mu\nu}, S_\nu]\Vert \sig'_\ka]}\cdot
  I^{-\ka}_{xx'}   -\sfrac14\Erw{i[F^{\mu\nu},
  D_\nu] \Vert\sig'_\ka]}\cdot I^{+\ka}_{xx'}  
\notag\\
 &&
 -\sfrac12\Erw{S^{\mu\nu}\Vert \sig_\nu}\cdot \delta_{xx'}-\sfrac13\Erw{F^{\mu\nu}\Vert
   i[\phi,i[\phi,D_\nu]]}\cdot \delta_{xx'}.
\eea
Notice the appearance of $\sig'_\nu\cdot I^{\pm\nu}_{xx'}$ within the
string-delta part $U_2\vert_{I\delta}$ of $U_2$. They are the $L$-$V$
analogues of $w_2(x,x')$ in $L$-$Q$, and will be treated as ``inert
fields'' in the following, according to ``Option (B)'' in
\sref{s:interlude}. However, we write them as they stand, by lack of
an $L$-$V$ version of \lref{l:aprio}, so that we will have to do all
cancellations by ``brute force''.

All obstruction maps are of the form 
  \bea{obst-F}
 O^\mu_{\erw{F_{\mu\nu}\Vert T^\nu }}(\phi') &=& -T_\nu\cdot
 iI^{+\nu}_{x'x},\notag\\
 O^\mu_{\erw{F_{\mu\nu}\Vert T^\nu }}(D'^\ka)  &=& T_\nu \cdot
 \pa^\ka iI^{+\nu}_{x'x},\notag\\
 O^\mu_{\erw{F_{\mu\nu}\Vert T^\nu }}(S'^\ka)  &=& 
T_\nu\cdot \pa^\ka iI^{-\nu}_{x'x}-2T^\ka \cdot \id ,\notag\\
O^\mu_{\erw{F_{\mu\nu}\Vert T^\nu }}(F'^{\ka\la})  &=&
 -T^{[\ka}\cdot \pa^{\la]}\id,\\
\label{obst-j}
 O^\mu_{\erw{j_\mu \Vert R}}(\Erw{T'_\ka\Vert j'^\ka}) &=&
 \Erw{i[R,T_\ka]\Vert j^\ka}\cdot\id, 
 \eea
 where $R$ and $T_\nu$ are Wick polynomials in $\phi$, $S$ and $D$
 only, which do not produce obstructions. They may be
 multi-local, depending on variables not indicated, and involving
 $\delta$ or $I\delta$.

We shall freely use \eref{cart-inv}, and
$\pa\delta_{xx'} = -\pa'\delta_{xx'}$ and $\pa I^\pm_{xx'} = -\pa' I^\pm_{xx'}$.

\paragraph{3. Vanishing of $L_3\vert_{\rm gl}$ and $K_3\vert_{\rm gl}$.}
We have to compute the purely gluonic parts of $P^{(3)}$ given by \eref{P3}. Because, on
account of \eref{Fjgen}, these arise exclusively through the gluonic
parts of \eref{LKU1} and \eref{LKU2}, they coincide with $P^{(3)}_{\rm
YM}$ of pure Yang-Mills. 

We first compute $O_{U_1'}(L_1-K_1)\vert_{\rm gl}= O_{U_1'\vert_{\rm gl}}((L_1-K_1)\vert_{\rm gl})$:
\bea{O11g} O_{U_1'}(L_1-K_1)\vert_{\rm gl}  &\!\!=\!\!& -\sfrac12\Erw{\sig'^{\ka} \Vert
  D_{\ka\la}}\cdot \pa'^{\la}\id  
+\sfrac14\Erw{F_{\ka\la}\Vert i[\sig'_\nu, D^\la]}\cdot \pa'^\ka iI^{-\nu}_{xx'}\notag \\
&&\quad
 -\sfrac12\Erw{F_{\ka\la}\Vert i[\sig'^\ka, D^\la]}\cdot \id +\sfrac14\Erw{F_{\ka\la}
   \Vert i[S^\ka,\sig'_\nu]}\cdot \pa'^\la iI^{+\nu}_{xx'}\notag \\
&\!\!=\!\!& \sfrac 12 \pa'^{\ka}\big(\Erw{\sig^{\la}\Vert
  D_{\ka\la}}\cdot\id\big) -\sfrac14 \pa^\ka\big(
\Erw{F_{\ka\la}\Vert i[\sig'_\nu, D^\la]} \cdot
iI^{-\nu}_{xx'}+\Erw{F_{\ka\la}\Vert i[\sig'_\nu,S^\la]}\cdot
iI^{+\nu}_{xx'}  \big)
\notag \\
&&-\sfrac 14
\Erw{\pa^{[\ka}\sig^{\la]}\Vert D_{\ka\la}}\cdot\id -\sfrac12 \Erw{F_{\ka\la}\Vert i[\sig^\ka,D^\la]}\cdot\id
.\quad
\eea
We have to apply $O_{U_1}$. Its action on the first derivative term in
\eref{O11g} is conveniently simplified by \lref{l:Opa}, where it produces two
derivative terms (which we are not interested in) and a bulk term
which vanishes because $\erw{\sigma^\la \Vert D_{\ka\la}}$ does not
contain $F$. The other derivative term, when evaluated via \lref{l:Opa},
produces a bulk term 
$-\sfrac14 O_Y(\pa''U_1''\vert_{\rm gl})$ with $Y^\mu =
\Erw{F^{\mu\rho}\Vert i[\sig'_\nu, D_\rho]} \cdot
iI^{-\nu}_{xx'}+\Erw{F^{\mu\rho}\Vert i[\sig'_\nu,S_\rho]}\cdot
iI^{+\nu}_{xx'}$, where everything except $F$ is ``inert'' by Option
(B). 
By \eref{pasig} and the Jacobi identity, the last line of \eref{O11g}
equals $-\sfrac i2
(\Erw{D^{\ka\la}\Vert D_{\ka\la}} -\Erw{F^{\ka\la}\Vert i[i[\phi,D_\ka],S_\la]})\cdot\delta_{xx'}$.
Thus, $O_{U_1''}(O_{U_1'}(L_1-K_1))\vert_{\rm
  gl}=O_{U_1''\vert_{\rm gl}}(O_{U_1'}(L_1-K_1)\vert_{\rm gl})$ yields
\bea{O111g}\sfrac12O_{U_1''}(O_{U_1'}(L_1-K_1))\vert_{\rm gl}
&\!\!\stackrel{\rm der}=\!\!&-\sfrac i{16} O^\mu_{\erw{F_{\mu\rho}
    \Vert i[\sig'_\nu, S^\rho]}}\big(\Erw{F''^{\ka\la}\Vert
  D''_{\ka\la}}\big)\cdot I^{+\nu}_{xx'}-\sfrac i{16}
O^\mu_{\erw{F_{\mu\rho} \Vert i[\sig'_\nu, D^\rho]}}\big(\Erw{F''^{\ka\la}\Vert D''_{\ka\la}}\big)  \cdot
I^{-\nu}_{xx'}
\notag\\ &&-\sfrac i8O^\mu_{\erw{F''_{\mu\nu}\Vert \sig''^\nu}}\big(\Erw{D^{\ka\la}\Vert D_{\ka\la}} -\Erw{F^{\ka\la}\Vert i[i[\phi,D_\ka],S_\la]}\big)\cdot\delta_{xx'}.\eea

The next term to compute is $O_{U_2}(L''_1+K''_1)\vert_{\rm
  gl}=O_{U_2\vert_{\rm gl}}((L''_1+K''_1)\vert_{\rm gl})$, where
\lref{l:delta} applies in the last term: 
\bea{O21g} -\sfrac{3i}2 O_{U_2}(L''_1+K''_1)\vert_{\rm gl} &\!\!=\!\!& -\sfrac{3i}{16}O^\mu_{\erw{F_{\mu\rho}\Vert i[\sig'_\nu,S^\rho]}}\big(\Erw{F''^{\ka\la}\Vert S''_{\ka\la}}\big)\cdot I^{-\nu}_{xx'} -\sfrac{3i}{16}O^\mu_{\erw{F_{\mu\rho}\Vert i[\sig'_\nu,D^\rho] }}\big(\Erw{F''^{\ka\la}\Vert S''_{\ka\la}}\big)\cdot
    I^{+\nu}_{xx'} 
\notag \\ &&  +\sfrac{i}8
 O^\mu_{\erw{F_{\mu\nu}\Vert i[\phi,i[\phi,D^\nu]]
   }}\big(\Erw{F''^{\ka\la}\Vert S''_{\ka\la}}\big) \cdot \delta_{xx'}.
\eea
The last term to compute is $O_{U'_1}(L_2+K_2)\vert_{\rm gl}=O_{U'_1\vert_{\rm gl}}(L_2+K_2)$:
\bea{O12g}-\sfrac{3i}2O_{U'_1}(L_2+K_2)\vert_{\rm gl}\cdot \delta_{xx''} &=&
\sfrac{3i}{16} O^\mu_{\erw{F'_{\mu\nu}\Vert \sig'^\nu}}\big(\Erw{S^{\ka\la}\Vert
  S_{\ka\la}}+\Erw{D^{\ka\la}\Vert D_{\ka\la}}\big) \cdot \delta_{xx''}.
\eea

\begin{prop}\label{p:LK3gl}
\bea{LK3g} L_3\vert_{\rm gl}=K_3\vert_{\rm gl}=0.\eea
\end{prop}

{\em Proof:} The proof is ``brute force'' and very lengthy with
enormous cancellations in the end. Especially paragraph 3.1 and 3.2
would probably be obsolete 
with an $L$-$V$ version of the apriori cancellation as in
\lref{l:aprio}.

The expressions \eref{O111g}, \eref{O21g}, \eref{O12g} can be worked out with
\eref{obst-F}. Their sum, modulo total derivatives, determines $(L_3-K_3)\vert_{\rm  glu}$.
We have to show that \eref{O111g} + \eref{O21g} + \eref{O12g} is a total
derivative. 

All terms involving two
string-integrated delta functions like $I^{\pm \nu}_{xx'}\cdot
\pa^\ka iI^{\pm\rho}_{x''x}$, arising only in the first
lines of \eref{O111g} and \eref{O21g}, are total derivatives (because
$\pa^\ka i[F_{\ka\la},S^\la] = \pa^\ka i[F_{\ka\la},D^\la]=0$) and can be ignored.
For the same reason, we can ignore the string-delta contribution to
the last term in \eref{O21g}.

To evaluate $O_{\erw{T\Vert F}}(S_{\ka\la})$ and $O_{\erw{T\Vert
    F}}(D_{\ka\la})$, we use
$S_{\ka\la}=\sfrac12(i[S_\ka,S_\la]+i[D_\ka,D_\la])$ and $D_{\ka\la}=
\sfrac12(i[S_\ka,D_\la]+i[D_\ka,S_\la])$. We discard derivatives, and permute $x'\lra x''$
when appropriate. We use re-bracketting  and the Jacobi identity. We collect
terms in \eref{O111g} + \eref{O21g} + \eref{O12g}:

\paragraph{3.1. Terms involving $I^{-\nu}$.} Either manifest or via 
$O_{\erw{T\Vert F}}(S)$.
\vskip-5mm
\bea{I-g} -\sfrac i{16}
\Erw{O^\mu_{\erw{F_{\mu\rho} \Vert i[\sig'_\nu, D^\rho]}}(F''^{\ka\la})\Vert D''_{\ka\la}} \cdot
I^{-\nu}_{xx'}
&\stackrel{\rm der}=& -\sfrac 1{8}  \pa''^\la \Erw{i[\sig'_\nu,
  D^\ka]\Vert D''_{\ka\la}} \cdot I^{-\nu}_{xx'}\delta_{xx''}
\notag \\
-\sfrac i{16}
\Erw{F''^{\ka\la}\Vert i[O^\mu_{\erw{F_{\mu\rho} \Vert i[\sig'_\nu, D^\rho]}}(S''_\ka)\vert_\delta,D''_\la]} \cdot
I^{-\nu}_{xx'}
&=&  -\sfrac 1{8} \Erw{F''^{\ka\la}\Vert i[i[\sig'_\nu,
  D_\ka],D''_\la]}  \cdot I^{-\nu}_{xx'}\delta_{xx''}
\notag\\
-\sfrac i4\Erw{D^{\ka\la}\Vert i[O^\mu_{\erw{F'_{\mu\nu}\Vert
    \sig'^\nu}}(S_\ka)\vert_{I\delta},D_\la]}\cdot\delta_{xx''}
&\stackrel{\rm der}=&  +\sfrac14 \pa_\ka \Erw{D^{\ka\la}\Vert
  i[\sig'_\nu,D_\la]} \cdot I^{-\nu}_{xx'}\delta_{xx''}
\notag   \\
+\sfrac i8\Erw{F^{\ka\la}\Vert
  i[i[\phi,D_\ka],O^\mu_{\erw{F'_{\mu\nu}\Vert
      \sig'^\nu}}(S_\la)\vert_{I\delta}]}\cdot\delta_{xx''}
&\stackrel{\rm der}=& -\sfrac18 \pa_\la \Erw{F^{\ka\la}\Vert
  i[i[\phi,D_\ka], \sig'_\nu]} \cdot I^{-\nu}_{xx'}\delta_{xx''}
\notag \eea
\bea{}
-\sfrac{3i}{16}\Erw{O^\mu_{\erw{F_{\mu\rho}\Vert
      i[\sig'_\nu,S^\rho]}}(F''^{\ka\la})\Vert S''_{\ka\la}}\cdot
I^{-\nu}_{xx'}
&\stackrel{\rm der}=&  -\sfrac3{8} \pa''^\la \Erw{
  i[\sig'_\nu,S^\ka]\Vert S''_{\ka\la}} \cdot
I^{-\nu}_{xx'}\delta_{xx''}
\notag \\
-\sfrac{3i}{16}\Erw{F''^{\ka\la}\Vert i[O^\mu_{\erw{F_{\mu\rho}\Vert
      i[\sig'_\nu,S^\rho]}}(S''_\ka)\vert_\delta,S''_\la]}\cdot I^{-\nu}_{xx'}
&=& -\sfrac3{8}\Erw{F''^{\ka\la}\Vert i[
  i[\sig'_\nu,S_\ka],S''_\la]} \cdot I^{-\nu}_{xx'}\delta_{xx''}
\notag \\
+\sfrac{i}8 \Erw{F''^{\ka\la}\Vert i[ O^\mu_{\erw{F_{\mu\nu}\Vert i[\phi,i[\phi,D^\nu]]
    }}(S''_\ka)\vert_{I\delta},S''_\la]} \cdot \delta_{xx'}
&\stackrel{\rm der}=& 0
\notag \\
+\sfrac{3i}{8} \Erw{S^{\ka\la}\Vert
  i[O^\mu_{\erw{F'_{\mu\nu}\Vert \sig'^\nu}}(S_\ka)\vert_{I\delta},S_\la]} \cdot
\delta_{xx''}
&\stackrel{\rm der}=& -\sfrac38 \pa_\ka \Erw{S^{\ka\la}\Vert
  i[\sig'_\nu,S_\la]} \cdot I^{-\nu}_{xx'}\delta_{xx''}
\notag \\
+\sfrac{3i}{8} \Erw{D^{\ka\la}\Vert i[O^\mu_{\erw{F'_{\mu\nu}\Vert \sig'^\nu}}(S_\ka)\vert_{I\delta},D_\la]} \cdot
\delta_{xx''}
&\stackrel{\rm der}=& -\sfrac38 \pa_\ka \Erw{D^{\ka\la}\Vert
  i[\sig'_\nu,D_\la]} \cdot I^{-\nu}_{xx'}\delta_{xx''}
\notag \eea

With the help of re-bracketting, we write the sum as $\frac18\Erw{X^- \Vert
  \sig_\nu'}\cdot I^{-\nu}_{xx'}\delta_{xx''}$ where
\bea{}
X^- &=& - i[D^\ka, \pa^\la D_{\ka\la}]
- i[i[F^{\ka\la}, D_\la],D_\ka] 
- 2\pa_\ka i[D^{\ka\la},D_\la]
- \pa_\la i[F^{\ka\la},  i[\phi,D_\ka]]
\notag \\ 
&& -3  i[S^\ka,\pa^\la S_{\ka\la}]
-3 i[i[F^{\ka\la},S_\la],S_\ka]
+3 \pa_\ka  i[S^{\ka\la},S_\la]
+3 \pa_\ka i[D^{\ka\la},D_\la]
\notag \\
&=& - i[\pa^\ka D_{\ka\la}, D^\la]
+ \sfrac 12 i[F^{\ka\la}, i[D_\ka,D_\la]] 
- 2i[\pa_\ka D^{\ka\la},D_\la]
+ i[F^{\ka\la},  i[D_\ka,D_\la]]
\notag \\ 
&&-3  i[\pa^\ka S_{\ka\la},S^\la]
+\sfrac32 i[F^{\ka\la},i[S_\ka,S_\la]]
+3  i[\pa_\ka S^{\ka\la},S_\la] -3  i[ F_{\ka\la},S^{\ka\la}]
+3 i[\pa_\ka D^{\ka\la},D_\la]
\notag \\
&=&\sfrac 12 i\big[F^{\ka\la}, i[D_\ka,D_\la]] +2 i[D_\ka,D_\la] 
+3 i[S_\ka,S_\la]
-6  S_{\ka\la}\big]=0.
\eea
\paragraph{3.2. Terms involving $I^{+\nu}$.}
Either manifest or via $O_{\erw{T\Vert F}}(D)$ or $O_{\erw{T\Vert F}}(\phi)$.
\vskip-5mm
\bea{I+g}  -\sfrac i{16} \Erw{O^\mu_{\erw{F_{\mu\rho} \Vert i[\sig'_\nu, S^\rho]}}(F''^{\ka\la})\Vert
  D''_{\ka\la}}\cdot I^{+\nu}_{xx'}
&\stackrel{\rm der}=&  -\sfrac1{8}\pa''^\la \Erw{i[\sig'_\nu, S^\ka]\Vert
  D''_{\ka\la}} \cdot I^{+\nu}_{xx'}\delta_{xx''}
\notag \\
-\sfrac i{16} \Erw{F''^{\ka\la}\Vert
  i[O^\mu_{\erw{F_{\mu\rho} \Vert i[\sig'_\nu,
      S^\rho]}}(S''_\ka)\vert_\delta,D''_\la]}\cdot I^{+\nu}_{xx'}
&=& -\sfrac 1{8} \Erw{F''^{\ka\la}\Vert
  i[i[\sig'_\nu,
  S_\ka],D''_\la]} \cdot I^{+\nu}_{xx'}\delta_{xx''}
\notag \\
-\sfrac i4\Erw{D^{\ka\la}\Vert i[S_\ka,O^\mu_{\erw{F'_{\mu\nu}\Vert
      \sig'^\nu}}(D_\la)]}\cdot\delta_{xx''}
&\stackrel{\rm der}=& +\sfrac14 \pa_\la \Erw{D^{\ka\la}\Vert i[S_\ka,
  \sig'^\nu]} \cdot I^{+\nu}_{xx'}\delta_{xx''}
\notag \\
+\sfrac i8\Erw{F^{\ka\la}\Vert i[i[O^\mu_{\erw{F'_{\mu\nu}\Vert
      \sig'^\nu}}(\phi),D_\ka],S_\la]}\cdot\delta_{xx''}
&=& +\sfrac18 \Erw{F^{\ka\la}\Vert i[i[\sig'^\nu,D_\ka],S_\la]} \cdot
I^{+\nu}_{xx'}\delta_{xx''}
\notag \\
+\sfrac i8\Erw{F^{\ka\la}\Vert i[i[\phi,O^\mu_{\erw{F'_{\mu\nu}\Vert
      \sig'^\nu}}(D_\ka)],S_\la]}\cdot\delta_{xx''}
&\stackrel{\rm der}=&  -\sfrac18 \pa_\ka \Erw{F^{\ka\la}\Vert
  i[i[\phi,\sig'^\nu],S_\la]} \cdot I^{+\nu}_{xx'}\delta_{xx''}
\notag \\
-\sfrac{3i}{16}\Erw{O^\mu_{\erw{F_{\mu\rho}\Vert i[\sig'_\nu,D^\rho]}}(F''^{\ka\la})\Vert S''_{\ka\la}}\cdot
I^{+\nu}_{xx'}
&\stackrel{\rm der}=& -\sfrac3{8}  \pa''^\la
\Erw{i[\sig'_\nu,D^\ka]\Vert S''_{\ka\la}}\cdot
I^{+\nu}_{xx'}\delta_{xx''}
\notag \\
-\sfrac{3i}{16}\Erw{F''^{\ka\la}\Vert i[O^\mu_{\erw{F_{\mu\rho}\Vert i[\sig'_\nu,D^\rho] }}(S''_\ka)\vert_\delta,S''_\la]}\cdot
I^{+\nu}_{xx'}
&=& -\sfrac3{8} \Erw{F''^{\ka\la}\Vert
  i[i[\sig'_\nu,D_\ka],S''_\la]} \cdot I^{+\nu}_{xx'}\delta_{xx''}
\notag \\
+\sfrac{i}8 \Erw{F''^{\ka\la}\Vert i[ O^\mu_{\erw{F_{\mu\nu}\Vert i[\phi,i[\phi,D^\nu]]
    }}(D''_\ka),D''_\la]} \cdot \delta_{xx'}
&\stackrel{\rm der}=& 0
\notag \\
+\sfrac{3i}{8} \Erw{S^{\ka\la}\Vert
  i[O^\mu_{\erw{F'_{\mu\nu}\Vert \sig'^\nu}}(D_\ka),D_\la]} 
\cdot \delta_{xx''}
&\stackrel{\rm der}=& -\sfrac38 \pa_\ka \Erw{S^{\ka\la}\Vert
  i[\sig'^\nu,D_\la]} \cdot I^{+\nu}_{xx'}\delta_{xx''}
\notag \\
+\sfrac{3i}{8} \Erw{D^{\ka\la}\Vert i[S_\ka,O^\mu_{\erw{F'_{\mu\nu}\Vert \sig'^\nu}}(D_\la)]} \cdot
\delta_{xx''}
&\stackrel{\rm der}=& -\sfrac38 \pa_\la \Erw{D^{\ka\la}\Vert
  i[S_\ka,\sig'^\nu]} \cdot I^{+\nu}_{xx'}\delta_{xx''}
\notag 
\eea
With the help of rebracketting, we write the sum as $\frac18\Erw{X^+\Vert
  \sig_\nu'}\cdot I^{+\nu}_{xx'}\delta_{xx''}$ where
\bea{}
X^+ &=& -i[S^\ka,\pa^\la D_{\ka\la}] 
-i[i[F^{\ka\la},D_\la],S_\ka] 
+2 \pa_\la i[D^{\ka\la},S_\ka]
+i[i[F^{\ka\la},S_\la],D_\ka] 
+i[i[F^{\ka\la},S_\la],D_\ka]
\notag \\
&&-3 i[D^\ka,\pa^\la S_{\ka\la}]
-3 i[i[F^{\ka\la},S_\la],D_\ka]
+3 \pa_\ka i[S^{\ka\la},D_\la] 
-3 \pa_\la i[D^{\ka\la},S_\ka]
-i[\pa^\ka D_{\ka\la},S^\la] 
\notag \eea
\bea{}
&=& -i[i[F^{\ka\la},D_\la],S_\ka] 
-2  i[\pa_\ka D^{\ka\la},S_\la]+2  i[F_{\ka\la},D^{\ka\la}]
+i[i[F^{\ka\la},S_\la],D_\ka] +i[i[F^{\ka\la},S_\la],D_\ka]
\notag \\
&&
-3 i[\pa^\ka S_{\ka\la},D^\la]
-3 i[i[F^{\ka\la},S_\la],D_\ka]
+3 i[\pa_\ka S^{\ka\la},D_\la] 
+3  i[\pa_\ka D^{\ka\la},S_\la] -3  i[F_{\ka\la},D^{\ka\la}] 
\notag \\
&=& -i[i[F^{\ka\la},D_\la],S_\ka]  -i[i[F^{\ka\la},S_\la],D_\ka] 
- i[F_{\ka\la},D^{\ka\la}] =0.
\notag
\eea
\paragraph{3.3. Terms involving $\delta_{xx'x''}$.}
\vskip-5mm
\bea{dtotg} -\sfrac i4\Erw{D^{\ka\la}\Vert i[O^\mu_{\erw{F''_{\mu\nu}\Vert
      \sig''^\nu}}(S_\ka)\vert_\delta,D_\la]}\cdot\delta_{xx'}
&=& -\sfrac12 \Erw{D^{\ka\la}\Vert i[\sig''_\ka,D_\la]}
\cdot \delta_{xx'x''}
\notag \\
+\sfrac i8\Erw{O^\mu_{\erw{F''_{\mu\nu}\Vert
      \sig''^\nu}}(F^{\ka\la})\Vert i[i[\phi,D_\ka],S_\la]}\cdot\delta_{xx'}
&\stackrel{\rm der}=& +\sfrac18 \Erw{\pa''^{[\ka}\sig''^{\la]}\Vert  i[i[\phi,D_\ka],S_\la]} \cdot \delta_{xx'x''} \notag \\
+\sfrac i8\Erw{F^{\ka\la}\Vert i[i[\phi,D_\ka],O^\mu_{\erw{F''_{\mu\nu}\Vert
      \sig''^\nu}}(S_\la)\vert_\delta]}\cdot\delta_{xx'}
&=& +\sfrac14 \Erw{F^{\ka\la}\Vert
  i[i[\phi,D_\ka],\sig''_\la]}\cdot \delta_{xx'x''}
\notag \\
+\sfrac i8\Erw{O^\mu_{\erw{F_{\mu\nu}\Vert i[\phi,i[\phi,D^\nu]]
    }}(F''^{\ka\la})\Vert S''_{\ka\la}} \cdot \delta_{xx'}
&\stackrel{\rm der}=& -\sfrac14 
\Erw{\pa^\la i[\phi,i[\phi,D^\ka]]\Vert S''_{\ka\la}}\cdot \delta_{xx'x''}
\notag \\
+\sfrac{i}8\Erw{F''^{\ka\la}\Vert i[ O^\mu_{\erw{F_{\mu\nu}\Vert i[\phi,i[\phi,D^\nu]]
    }}(S''_\ka)\vert_\delta,S''_\la]} \cdot \delta_{xx'}
&=& +\sfrac14 \Erw{F''^{\ka\la}\Vert
  i[i[\phi,i[\phi,D_\ka]],S''_\la]}\cdot \delta_{xx'x''}
\notag \\
+\sfrac{3i}{8} \Erw{S^{\ka\la}\Vert
  i[O^\mu_{\erw{F'_{\mu\nu}\Vert \sig'^\nu}}(S_\ka)\vert_\delta,S_\la]} \cdot
\delta_{xx''}
&=& +\sfrac34 \Erw{S^{\ka\la}\Vert
  i[\sig'_\ka,S_\la]}\cdot \delta_{xx'x''}
\notag \\
+\sfrac{3i}{8} \Erw{D^{\ka\la}\Vert i[O^\mu_{\erw{F'_{\mu\nu}\Vert \sig'^\nu}}(S_\ka)\vert_\delta,D_\la]} \cdot
\delta_{xx''}
&=& +\sfrac34  \Erw{D^{\ka\la}\Vert
  i[\sig'_\ka,D_\la]} \cdot \delta_{xx'x''}
\notag 
\eea
The sum is $\frac14 Z\cdot \delta_{xx'x''}$ where
\bea{}
Z&=&-2 \Erw{D^{\ka\la}\Vert i[\sig_\ka,D_\la]} 
+\Erw{D^{\ka\la}\Vert  i[i[\phi,D_\ka],S_\la]}
+\Erw{i[\phi,F^{\ka\la}]\Vert i[i[\phi,D_\ka],S_\la]}
\notag \\
&&+\Erw{F^{\ka\la}\Vert  i[i[\phi,D_\ka],\sig_\la]}
-\Erw{ i[D^\la,i[\phi,D^\ka]]\Vert S_{\ka\la}} -\Erw{i[\phi,i[D^\la,D^\ka]]\Vert S_{\ka\la}}
\notag \\
&&+\Erw{F^{\ka\la}\Vert  i[i[\phi,i[\phi,D_\ka]],S_\la]}
+3 \Erw{S^{\ka\la}\Vert  i[\sig_\ka,S_\la]}
+3 \Erw{D^{\ka\la}\Vert  i[\sig_\ka,D_\la]} .
\notag \eea
Terms involving $F$ vanish by Jacobi:
\bea{}
\Erw{F^{\ka\la}\Vert - i[\phi,
  i[i[\phi,D_\ka],S_\la]]+i[i[\phi,D_\ka],i[\phi,S_\la]]+
  i[i[\phi,i[\phi,D_\ka]],S_\la]} =0.
\notag
\eea
The remaining terms also vanish by Jacobi: 
\bea{}
Z&=& \Erw{D^{\ka\la}\Vert i[i[\phi,S_\ka],D_\la] 
+  i[i[\phi,D_\ka],S_\la]} +\Erw{S_{\ka\la}\Vert - i[D^\la,i[\phi,D^\ka]]
  -i[\phi,i[D^\la,D^\ka]] +3 i[i[\phi,S^\ka],S^\la]} 
\notag \\
&=& \Erw{D^{\ka\la}\Vert i[\phi,i[S_\ka,D_\la]} +\sfrac32\Erw{S_{\ka\la}\Vert
  i[\phi,i[D^\ka,D^\la]+i[S^\ka,S^\la]]} = \Erw{D^{\ka\la}\Vert
  i[\phi,D_{\ka\la}]}+ 3\Erw{S^{\ka\la}\Vert i[\phi,S_{\ka\la}]}= 0.
\notag
\eea
This completes the proof of \pref{p:LK3gl}. \qed

\paragraph{4. Vanishing of $L_3\vert_{\rm quark}$ and $K_3\vert_{\rm quark}$.}
We have to compute the quark parts of \eref{P3}. 
We shall compute the ``bulk'' terms (contributing to $(L_3-K_3)\vert_{\rm
qu}$) and the derivative terms (contributing to $U_3\vert_{\rm qu}$) separately.

We first compute $O_{U_1'}(L_1-K_1)\vert_{\rm qu}=
(O_{U_1'\vert_{\rm gl}}+O_{U_1'\vert_{\rm qu}})(\erw{D_\ka\Vert j^\ka})$:
\bea{O11q}
O_{U_1'}(L_1-K_1)\vert_{\rm qu} &=& -\sfrac i2 \pa^\ka W_\ka^+ + 
 \Erw{i[\phi,D_\ka] \Vert j^\ka}\cdot \id,
\eea
where we abbreviate
\bea{W} W^\pm_\ka := \Erw{\sig'_\nu\Vert j_\ka }\cdot I^{\pm\nu}_{xx'}.\eea

We have to apply $O_{U''_1}$ to \eref{O11q}. For the contribution of
the derivative term, we use \lref{l:Opa}. In the contribution of the bulk term, we exchange $x'\lra
x''$. Because, on account of \eref{Fjgen}, $O_{W^+}$ acts trivially on gluonic fields, this yields 
\bea{O111q} \sfrac12 O_{U''_1}(O_{U'_1}(L_1-K_1))\vert_{\rm qu} &=& -\sfrac
i4\pa^\ka O_{U''_1}(W^+_\ka) +\sfrac i4\pa''_\mu
O_{W^+}(\erw{\phi''\Vert j''^\mu}) -\sfrac i4O_{W^+}(\erw{D''_\ka\Vert j''^\ka}) \notag \\
&& -\sfrac14 (\pa^\ka+\pa''^\ka) \big(\Erw{
  i[\phi,\sig'_\nu] \Vert j_\ka}\cdot iI^{+\nu}_{xx'}\cdot
i\delta_{xx''}\big)
-\sfrac12\Erw{i[\sig'_\nu, D_\ka]\Vert j^\ka} \cdot
iI^{+\nu}_{xx'}\cdot i\delta_{xx''}
  \notag \\ 
&&+\sfrac 12 \Erw{i[\phi,  i[\phi,D_\ka]] \Vert j^\ka}\cdot
i\delta_{xx'}\cdot i\delta_{xx''}.
\eea

We next compute $O_{U_2}(L''_1+K''_1)\vert_{\rm qu}=(O_{U_2\vert_{\rm gl}}+O_{U_2\vert_{\rm qu}})(\erw{S''_\ka\Vert j''^\ka})$. Notice that $U_2\vert_{\rm qu}=\frac12W^-$. Thus,
\bea{O21q}-\sfrac{3i}2  O_{U_2}((L_1''+K_1''))\vert_{\rm qu} &=&
\sfrac38
\big(\Erw{i[S_\nu,\sig'_\ka]\Vert j''_\la} \cdot iI^{-\ka}_{xx'}
+\Erw{i[D_\nu,\sig'_\ka]\Vert j''_\la} \cdot iI^{+\ka}_{xx'} \big)\cdot
(-2\eta^{\nu\la}i\delta_{xx''} -\pa''^\la iI^{-\nu}_{x''x})
\notag \\
&&+\sfrac14 \Erw{i[\phi,i[\phi,D_\nu]]\Vert j''_\ka}
\cdot (-2\eta^{\nu\ka}i\delta_{xx''} -\pa''^\ka iI^{-\nu}_{x''x})\cdot\id 
\notag \\ && -\sfrac {3i}4O_{W^-}(\erw{S''_\ka\Vert j''^\ka}),\eea
where the two terms involving $\pa'' I^-_{x''x}$ are total derivatives
because $j''$ is conserved.

Finally, $O_{U_2}(L_1+K_1)\vert_{\rm qu}=0$.

Now, we have in \eref{O111q} and in \eref{O21q}
\bea{} O_{W^+}(\Erw{D''_\ka\Vert j''^\nu})=\Erw{i[\sig'_\nu,D_\ka] \Vert
  j^\ka}\cdot i\delta_{xx''}\cdot I^{+\nu}_{xx'} ,\quad
O_{W^-}(\Erw{S''_\ka\Vert j''^\nu})=\Erw{i[\sig'_\nu,S_\ka] \Vert
  j^\ka}\cdot i\delta_{xx''}\cdot I^{-\nu}_{xx'} .\eea

One then sees directly that the bulk
terms from \eref{O111q} and \eref{O21q} cancel each other. Thus, we
have proven

\begin{prop}\label{LK3qu}
\bea{LK3q}L_3\vert_{\rm qu}=K_3\vert_{\rm qu}=0.\eea
\end{prop}

\paragraph{5. Computation of $U_3\vert_{\rm quark}$.}

We collect the derivative terms from \eref{O111q} and
\eref{O21q}. Appropriately permuting the arguments $x\lra x''$, we
read off by \eref{P3}:
\bea{U3qraw} U_3^\mu\vert_{\rm qu}(x;x',x'') &=&
\sfrac i4O_{U''_1}(\Erw{\sig'_\nu\Vert
  j^\mu})\cdot
  I^{+\nu}_{xx'}-\sfrac i4O_{\erw{\sig'_\nu\Vert j''}}(\erw{\phi\Vert j^\mu}) \cdot
  I^{+\nu}_{x''x'}- \sfrac12 \Erw{
  i[\phi,\sig'_\nu] \Vert j^\mu }\cdot I^{+\nu}_{xx'}\cdot
\delta_{xx''} \notag \\
&&-\sfrac38 \big(\Erw{i[S''_\nu,\sig'_\ka]\Vert j^\mu} \cdot I^{-\ka}_{x''x'}
+\Erw{i[D''_\nu,\sig'_\ka]\Vert j^\mu} \cdot I^{+\ka}_{x''x'} \big)\cdot
I^{-\nu}_{xx''}\notag \\
&&
-\sfrac14 \Erw{i[\phi'',i[\phi'',D''_\nu]]\Vert j^\mu}\cdot 
I^{-\nu}_{xx''}\cdot \delta_{x'x''}.
\eea
The first term of \eref{U3qraw} contains
\bea{first} \sfrac i4 O_{U_1''\vert_{\rm gl}}(\Erw{\sig'_\nu\Vert   j^\mu})\cdot
  I^{+\nu}_{xx'}&=&\sfrac i8 \big(\Erw{
  -(i[\sig''_\ka ,S'_\nu]\Vert   j^\mu}\cdot
  iI^{+\ka}_{x'x''}+\Erw{i[\phi',\sig''_\ka]\Vert   j^\mu}\cdot(-2\eta^{\ka\nu}i\delta_{x'x''} +
  \pa''_\nu iI^{-\ka}_{x'x''})\big)\cdot
  I^{+\nu}_{xx'} \notag \\
&=&\sfrac 18 \Erw{i[\sig''_\ka ,S'_\nu]\Vert   j^\mu}\cdot
  I^{+\ka}_{x'x''}\cdot   I^{+\nu}_{xx'}\notag \\ &&+\sfrac
  14\Erw{i[\phi',\sig''_\ka]\Vert   j^\mu}\cdot \delta_{x'x''}\cdot   I^{+\ka}_{xx'}  +\sfrac 18\Erw{i[\phi',\sig''_\ka]\Vert   j^\mu}\cdot
  \pa'_\nu I^{-\ka}_{x'x''}\cdot   I^{+\nu}_{xx'},
 \eea
 while the remaining terms in the first line of \eref{U3qraw} sum up
 to $-\erw{i[\phi,\sig_\nu']\Vert j^\mu}\cdot I^{+\nu}_{xx'} \cdot
 \delta_{xx''}$ by \eref{obst-j}. Thus, we have proven

\begin{prop}\label{U3qu}
 \bea{U3q} U_3^\mu\vert_{\rm qu}(x;x',x'') =  \Erw{\xi_3(x;x',x'')\Vert j^\mu(x)}
 \eea
with the purely gluonic field
\bea{xi3} \xi_3(x;x',x'')&=&\Sy_{x'x''}\Big[\sfrac 18 i[\sig''_\ka ,S'_\nu]\cdot
  I^{+\ka}_{x'x''}\cdot   I^{+\nu}_{xx'} +\sfrac
  14i[\phi',\sig'_\ka]\cdot I^{+\ka}_{xx'}\cdot \delta_{x'x''} +\sfrac 18i[\phi',\sig''_\ka]\cdot
\pa'_\nu I^{-\ka}_{x'x''}\cdot I^{+\nu}_{xx'}
 \notag \\ && -i[\phi,\sig_\nu']\cdot I^{+\nu}_{xx'} \cdot \delta_{xx''}
\notag\\ &&+ \sfrac38 \big(i[\sig'_\ka,S''_\nu]\cdot I^{-\ka}_{x''x'}+ 
i[\sig'_\ka,D''_\nu]\cdot I^{+\ka}_{x''x'}\big)\cdot
I^{-\nu}_{xx''}-\sfrac14 i[\phi',i[\phi',D'_\nu]]\cdot I^{-\nu}_{xx'}\cdot
\delta_{x'x''}\Big].\quad\eea
\end{prop}

\begin{conj}\label{cj:Unq}
We conjecture that the general structure \eref{U3q} holds for all
$U_n\vert_{\rm qu}$ with the gluonic fields $\xi_n$ only containing
$\phi$, $D$ and $S$.
\end{conj}

(It should be possible to prove the conjecture by induction.)

Because of \eref{U3q} and \eref{Ojj}, we have (certainly for $n=1,2,3$)
\bea{Unpsi}O_{U_n}(\psi(x)) = \wt \xi(y;y',\dots,y^{(n-1)}) j(x)\cdot \delta_{xy},\eea
where from \eref{LQV-QCDmin} and \eref{LKU2-QCD}
\bea{wtxi} \wt\xi_1(y)=\xi_1(y) = \phi(y),\qquad
\wt\xi_2(y;y')=\xi_2(y;y') = \sfrac12 \sig'_\nu \cdot
I^{-\nu}_{yy'},\eea
while $\wt\xi_3(y;y',y'')$ differs from $\xi_3(y;y',y'')$
in \eref{xi3} by a factor $\frac12$ on the term in the second line involving the factor
$\delta_{yy'}$ or $\delta_{yy''}$, because of \lref{l:delta}.

\subsection{The dressed quark field $\psi_{[g]}$}
\label{s:dressQCD}
Let
$$\Omega_{n}(X(x)):= \int dy\, dy' \dots dy^{(n-1)}\,
O_{U_n(y;y',\dots,y^{(n-1)})}(X(x)).$$
(One must be careful to carry out the integrals over $O_{U_n}$, rather
than over $U_n$, because of \lref{l:delta}.) The dressed quark
field is \cite{LV}
\bea{psig}\psi_{[g]}(x)=\exp (i\Omega_U)(\psi(x)),\eea
where
$$\Omega_{U}:= g\Omega_{1}+\sfrac{g^2}2\Omega_{2}+
\sfrac{g^3}6\Omega_{3}+\dots .$$

With \eref{Unpsi}, 
$$\Omega_{n}(\psi)(x) = \phi_n(x)\cdot\psi(x),$$
where
$$\phi_n(x):= \int dy\, dy'\dots\int  dy^{(n-1)}\, \wt\xi_n(y;y',\dots,y^{(n-1)})\cdot
\delta_{yx} = \int dy'\dots\int  dy^{(n-1)}\,
\wt\xi_n(x;y',\dots,y^{(n-1)}),  $$

Integrals over string-integrated $\delta$ functions will be carried out as
\bea{intIpm} \int dy'\, Y_\nu(y') \cdot I^{\pm \nu}_{yy'}=\int dy'\,
Y_\nu(y') \cdot (I^\nu\pm I_{-u}^\nu)(\delta)(y-y') = (I^\nu\pm
I_{-u}^\nu)(Y)(y)=: I^{\pm \nu}(Y)(y).\eea
This gives 
\bea{phi12}\phi_1=\phi, \qquad \phi_2 =\sfrac12 I^- (i[\phi,S]),\eea
\bea{phi3} \phi_3&=& -\sfrac 14i[\phi,I^-(i[\phi,S])] +\sfrac14
I^+(i[\phi,i[\phi,S]])+\sfrac18I^+\big(i[I^+(i[\phi,S]),S] + i[I^-(i[\phi,S]),D]\big)\notag\\ && -\sfrac 12i[\phi,I^+(i[\phi,S])] -\sfrac14
I^-(i[\phi,i[\phi,D]]) + \sfrac38 I^-\big(i[I^+(i[\phi,S]),D] +
i[I^-(i[\phi,S]),S]\big).\eea

We write
\bea{Wpsi}\psi_{[g]} = W(c)\cdot \psi\eea
with a string-dependent gluonic field $W(c)$ to be computed. 
Expanding \eref{psig} until the third order yields
\bea{W123exp} W(c)&=& 1 +ig \phi + \sfrac{(ig)^2}2\big(\phi^2 +
\Omega_1(\phi)-i\phi_2\big) \\ \notag
&&+\sfrac{(ig)^3}6(\phi^3+2\Omega_1(\phi)\phi+\phi \Omega_1(\phi) +
\Omega_1^2(\phi)-\sfrac
{3i}2\big(\phi_2\phi+\phi\phi_2+\Omega_1(\phi_2)+\Omega_2(\phi_1)\big)-\phi_3)
+ \dots
.\eea
When  \eref{phi12}, \eref{phi3}, and
\bea{Omegaphi} \Omega_1(\phi)&=&-\sfrac i2 I^+(i[\phi,S]), \notag \\
\Omega_1^2(\phi)&=&
-\sfrac 14
 I^+\big(i[I^+(i[\phi,S]),S]+i[I^-(i[\phi,S]),D]\big) - \sfrac 12
 I^+(i[\phi,i[\phi,S]]) + \sfrac12 i[\phi,I^-(i[\phi,S])], \notag \\
 \Omega_2(\phi)&=&-\sfrac i4
 I^+\big(i[I^-(i[\phi,S]),S]+i[I^+(i[\phi,S]),D]\big) + \sfrac i6
 I^+(i[\phi,i[\phi,D]]) , \notag \\
 \Omega_1(\phi_2)&=&-\sfrac
 i4I^-\big(i[I^+(i[\phi,S]),S]+i[I^-(i[\phi,D]),S]\big) -\sfrac i2 I^-(i[\phi,i[\phi,S]])
\eea
are inserted, all contributions involving $I_u$ cancel (via $I^++I^-=2I_c$):
\bea{W123comp} W(c)&=& 1 +ig \phi +
\sfrac{(ig)^2}2\big(\phi^2 -iI_c(i[\phi,S])\big) \notag \\  &&
+\sfrac{(ig)^3}6\big(\phi^3-3iI_c(i[\phi,S])\phi
-\sfrac12I_c(i[\phi,i[\phi,3S-D]]) - \sfrac32 I_c(i[I_c(i[\phi,S]),S+D])\big) + \dots\quad
\\ \label{W123expo} &=& \exp ig \Big[\phi +
\sfrac{g}2I_c(i[\phi,S]) \notag \\ &&+ \sfrac{g^2}6
\big(-\sfrac32i[\phi,I_c(i[\phi,S])] + \sfrac12 I_c(i[\phi,i[\phi,3S-D]])
+ \sfrac32 I_c(i[I_c(i[\phi,S]),S+D])\big)\dots\Big].
\eea
The third-order term of the expansion \eref{W123comp} coincides with
the expression 
\bea{IH} \phi^3 +3I_c([\phi,A(c)+A])\phi +I_c([\phi,[\phi,A(c)+2A]]) +3
I_c([I_c([\phi,A(c)+A]), A(c)]),
\eea
(first computed by I. Hemprich \cite{Hemp23}).
More precisely, \eref{IH} was derived for a {\em classical} gauge
potential by demanding that
$W(x,c)$ transforms like 
\bea{GTW} W(x,c)\stackrel!\mapsto W(x,c) e^{-i\lambda(x)},\eea
so as to make $\pi(W(x,c)) \psi(x)$ gauge invariant, under
infinitesimal non-abelian gauge transformation law 
($\lambda(x)$ Lie-algebra valued with $\lambda(x)\to0$ as $x\to\infty$)
\bea{GT}A_\mu(x) \mapsto A_\mu(x) + \pa_\mu\lambda(x) +g \cdot
[\lambda(x),A_\mu(x)].\eea
\eref{GT} entails the transformation of string-localized fields $\phi(x,c) = (I_cA)(x)$ and $A_\mu(x,c):= A_\mu(x) +
\pa_\mu \phi(x,c)$
\bea{}\phi(c) &\mapsto& \phi(c) - \lambda(x) + g \cdot
I_c(i[\lambda,A])(x), \notag\\ 
A_\mu(x,c) &\mapsto& A_\mu(x,c) + \pa_\mu\lambda(x) + g\cdot
\big(i[\lambda(x),A_\mu(x)] + \pa_\mu I_c(i[\lambda,A])(x)\big).\notag
\eea
\eref{W123comp} are the three lowest
orders satisfying \eref{GTW}. Moreover, they
specialize to the proper path-ordering along a string $e$ when the string smearing $c(e)$ is sharp.

In contrast, in sQFT, \eref{W123comp} is derived for normal-ordered
products of {\em quantum} fields $A\equiv A^u$ and
$A(c)\equiv A^u(c)$ on the Hilbert space $\HH^u$.

However, because each $A^u$ or $A^u(c)$ in
\eref{W123comp} is contracted
with $I_c$ and $(eu)=0$ for all $e$ in the support
of $c(e)$, \eref{W123comp} does not change when $A^u$ and $A^u(c)$ are
replaced by $A$ and $A(c)$.

Thus, at least until third order, $W(c)$ is a ``smeared
Wilson-operator'', sharing the same infinitesimal non-abelian gauge
transformation law \eref{GTW} as Wilson operators along any curve from $x$ to $\infty$.

It is group-valued (unitary when evaluated in any unitary
representation of the gauge group is a unitary group) because the exponent in \eref{W123expo} is $ig$ times a real
(hermitean) Lie-algebra-valued field.

\paragraph{The issue of $I_u$.} The systematic cancellation of all terms involving $I_{-u}$ (arising
from casting $I_u$, acting on the string-delta's, onto the fields as
in \eref{intIpm}) in
\eref{W123comp} suggests that one may omit $I_{u}$ from the outset,
i.e., already in Table 2 and in \eref{obst-F}. It should, however,
  be objected that $I_{-u}$ can contribute to the $I_{-u}$-independent result via
$(I_{-u}\pa)=-1$. Indeed, $\Omega_1^2(\phi)$, $\Omega_1(\phi_2)$ and
$\phi_3$ contain $(I_{-u}\pa)
i[\phi, I^-(\sig)]=-i[\phi, I^-(\sig)]$ with coefficients $-\frac14$, $+\frac i4$, and
$+\frac 18$, respectively. That these contributions separately cancel
in \eref{W123exp}, supports the suggestion.

\subsection{``Kinematic confinement?''}
\label{s:kinc}
B. Schroer has speculated whether the infrared behaviour of the dressed
quark field might already show signals of confinement, e.g., by one of
the criteria prevailing in the literature. 

Recall that the IR superselection structure of QED arises by using the Weyl algebra:
correlation functions of dressing factors are exponentials of the
two-point function of the escort field. The IR structure cannot be seen in any
finite order of the exponential series.

We expect the same also in the non-abelian case. Unfortunately, our
methods (both the obstruction formalism to compute the dressed field
in $L$-$V$, and the computation of the classical ``smeared Wilson
operator'' by imposing \eref{GTW}) are perturbative (recursive
order by order), and therefore not suitable for any control of
``kinematic confinement''. 

Some all-order results have been obtained by M. Pabst
\cite{Pab24} for the field
$$\pi(e^{ig\phi})\psi$$
with the ordinary exponential in the representation $\pi$ in which the
quarks transform. This field arises as the dressed field when the
initial $L$-$V$ pair is given by $K_1=0$
and $L_1 = \pa_\mu U_1^\mu =\pa_\mu \erw{\phi\Vert j^\mu}$ (notes KHR
2022/23: one finds $K_n=0$ in all orders and $L_n = \erw{\ad_{i\phi}^{n-1}(\pa
\phi)\Vert j^\mu}$ which amounts to a non-abelian pure-gauge vector potential).
The computation of its two-point function requires the knowledge of
the values of infinitely many Casimir operators
$$C^\pi_n := \sum_{\sig\in S_n}C^\pi_n(\sig), \qquad C^\pi_n(\sig) := \pi\big(\tau_{a_1}\dots \tau_{a_n}
\tau_{a_{\sig(n)}}\dots \tau_{a_{\sig(1)}}\big)$$
($n\in \NN$) in the representation $\pi$. These are not know (to me)
in closed form.

Pabst succeeded to compute the two-point function by a different
method, for the
spin $\frac12$ and spin $1$ representations of $su(2)$. In both cases, he found
essentially the same IR behaviour that leads to an IR superselection
structure, as in QED.

Thus, if there is any ``kinematic confinement'', then it cannot be
seen in this simple model. Starting from $K_1=\erw{A_\mu\Vert j^\mu}$
and $L_1 = \erw{A_\mu(c)\Vert j^\mu}$ (``colored QCD'') turns out to
be inconsistent (the SI condition for the S-matrix cannot be
fulfilled), unless one adds the $L$-$V$ pair of YM
(``lock-key''). Thus, one is forced to study the dressed field of QCD, as in
\sref{s:dressQCD}, with a non-perturbative closed expression for $W(c)$
lacking. 

Trying to compute the two-point function in lowest orders, one notices
two things. (i): Rather than the Casimir operators $C_n^\pi$, one needs all
Casimir operators $C_n^\pi(\sig)$ separately, and (ii): Without
smearing with $c(e)$, the resulting momentum integrals are not only
divergent (as in the abelian case) but ill-defined. In the
abelian case, the ``non-perturbative'' Weyl-algebra 
methods allowed to get a finite result.


\section{Higgs-Kibble}
\label{s:HK}

\paragraph{General strategy.} Let $B_\mu$ the local Proca field
$B_\mu(x)$ describing a free massive vector boson of mass $m$ on the Wigner Hilbert space of the
massive spin-1 representation, and $F_{\mu\nu}$ its field tensor, 
satisfying the Klein-Gordon equation and the Proca equations of motion
\bea{eomB},\quad
\pa_\mu B^\mu=0, \quad \pa_\mu B_\nu - \pa_\nu
B_\mu = F_{\mu\nu}, \quad \pa_\mu F^{\mu\nu} = -m^2 B^\mu.
\eea
$B$ has dimension scaling 2, so that minimal couplings are
beyond the power-counting bound. One therefore introduces
\bea{ABphi} 
A_{\mu}(x,c) &:=& (I^\nu_c F_{\mu\nu})(x) = B_\mu + \pa \phi_\mu(x,c), \\ \notag
\phi(x,c) &:=& (I^\mu_c B^\nu)(x) ,
\eea
both of dimension 1, so that minimal couplings of $A(c)$ are
power-counting renormalizable. Unlike
in QED, the field $\phi$ is defined on the Wigner Hilbert space
without any extension.

These fields also satisfy the Klein-Gordon equation, and the equations of motion
\bea{eomA} \pa_\mu A^\mu(c)=-m^2 \phi(c), \quad \pa_\mu\phi(c) = A_\mu(c)-B_\mu, \quad \pa_\mu A_\nu(c) - \pa_\nu
A_\mu(c) = F_{\mu\nu} .\eea

We are going to study the Higgs-Kibble model with $N$ massive vector
bosons, minimally coupled to fermionic vector and axial currents, self-coupled by a
massive variant of YM, and coupled to a scalar field describing a ``Higgs'' particle  of
mass $m_H$, all on their Wigner Hilbert spaces. 

The aim is to establish string-independence of the S-matrix (SI). In
particular, we want to see the necessity of the
Higgs coupling. We assume massless fermions so that $j^5_a$ are
conserved currents.

For a first overview, we start with the field content structure of the initial LV
pairs, given below: let ``$P$'' stand for any field in the Proca sector
($A,B,F,\phi,w$),
``$J$'' for lepton currents, and ``$H$'' for the Higgs field and its
derivative. Then
$$K_{1,\rm self} \quad\hbox{and}\quad L_{1,\rm self} \quad\hbox{and} \quad V_{1,\rm self} = P^3,$$
$$K_{1,\rm Higgs} \quad\hbox{and} \quad L_{1,\rm Higgs} \quad\hbox{and} \quad V_{1,\rm Higgs} = P^2 H + H^3,$$
$$K_{1,\rm min} \quad\hbox{and} \quad L_{1,\rm min} \quad\hbox{and} \quad V_{1,\rm min} = PJ$$
leads to the following table for the second-order obstructions:

$$\begin{tabular}{l||c|c|c|}
$O_{V_1}(K_1+L_1)$&minimal&self&Higgs\cr \hline\hline
$V_{1,\rm min}$&$J^2+\red{P^2J}$&$\red{P^2J}$&$PJH$\cr\hline
$V_{1,\rm self}$&$\red{P^2J}$&$\green{P^4}$&$P^3H$\cr\hline
$V_{1,\rm Higgs}$&$PJH$&$P^3H$&$\green{P^4}+P^2H^2$\cr\hline
\end{tabular}$$
{\bf Table 3.} Field content and cancellation patterns.
\bigskip

The necessary cancellation of non-resolvable obstructions marked in
red in the table means that non-abelian minimal coupling is inconsistent without
self-interaction, hence need also Higgs. This is an ordinary
``lock-key situation''.

The necessary cancellation of non-resolvable obstructions marked in
blue in the table means that non-abelian self-interacting massive
vector bosons are inconsistent
without Higgs, as claimed previously by JM, BS, JGB, and conversely,
the non-abelian Higgs coupling is also inconsistent
without self-coupling. This is a ``lock-key situation in both
directions''.

In particular, minimal interactions require both self-couplings and Higgs.

In gauge theory, different symmetry breaking patterns of different
gauge groups lead to different numbers of Higgs particles, different
mass ratios, and different numerical coefficients. In order to avoid a most
general ansatz with parameters that would allow all these solutions,
I test the above in the simple case of $N$ massive vector bosons with
equal masses, and one Higgs. 

Notations of Lie-algebra valued fields $X=X_a\tau_a$ is as in \sref{s:QCD}.
We do not assume $N=3$, nor $\mathfrak{su}(2)$ with $\tau_a=\frac12\sigma_a$, but
shall derive it from second-order SI.

\subsection{$L$-$Q$ and $L$-$V$ pairs.}

Throughout the rest of this section, we shall write $A(c)\equiv A$,
$\phi\equiv \phi(c)$ and $L_1\equiv L_1$, $V_1 \equiv V_1(c)$ and
$I\equiv I_c$ etc.\ because there is no ambiguity.

Let the fermions (electrons) have mass $m_e$. The vector and axial currents are $j^\mu_a = \ol\psi\gamma^\mu \tau_a\psi$ and
$j^{\mu5}_a=\ol\psi\gamma^\mu\gamma^5\tau_a\psi$. The vector currents are
conserved, while the axial currents are conserved only in the massless case:
\bea{}\pa_\mu j^{\mu5}_a=2m_e\cdot
S^5_a:=2m_e\cdot\ol\psi\gamma^5\tau_a\psi.\eea
We shall actually work with $m_e=0$. We set out with an arbitrary linear
combination
\bea{J} J^\mu = cj^\mu + c_5 j^{\mu5}.\eea

We have the $L$-$Q$ and $L$-$V$ pairs 
\bea{iniLVmin}
L_{1,\rm min} &=& \Erw{A_\mu \Vert J^\mu} +2m_e c_5\Erw{\phi\Vert S^5}, \notag\\
Q^\mu_{1,\rm min} &=& \Erw{w \Vert J^\mu},\notag\\
K_{1,\rm min} &=& \Erw{B_\mu \Vert J^\mu}, \notag\\
V^\mu_{1,\rm min} &=& \Erw{\phi \Vert J^\mu},
\eea
\bea{iniLVself}
L_{1,\rm self} &=& \sfrac 12 \Erw{F^{\mu\nu} \Vert 
i[A_\mu, A_\nu]} -\sfrac 12 m^2\Erw{\phi\Vert i[A^{\mu}, B_\mu] }, \notag\\
Q^\mu_{1,\rm self} &=& \Erw {F^{\mu\nu}\Vert i[w,A_\nu]} - \sfrac12
  m^2 \Erw{\phi\Vert i[w, B^\mu]},\notag\\
K_{1,\rm self} &=& \sfrac 12 \Erw{F^{\mu\nu} \Vert i[B_\mu, B_\nu]},\notag\\
V^\mu_{1,\rm self} &=& \sfrac 12 \Erw{F^{\mu\nu} \Vert i[\phi,A_\nu+B_\nu]},
\eea
\bea{iniLVhiggs}
L_{1,\rm Higgs}
&=&\mu\cdot\big(\Erw{A_\mu \Vert B^\mu}H+\Erw{A_\mu \Vert
  \phi}\pa^\mu H-\sfrac{m_H^2}2\Erw{\phi \Vert \phi}H+aH^3\big),
\notag\\
Q^\mu_{1,\rm Higgs} &=& \mu\cdot\big(\Erw{w\Vert B^\mu}H + \Erw{w\Vert \phi}\pa^\mu H\big) ,\notag\\
K_{1,\rm Higgs} &=& \mu\cdot\big(\Erw{B_\mu \Vert B^\mu}H+aH^3\big), \notag\\
V^\mu_{1,\rm Higgs} &=& \mu\cdot\big(\Erw{B^\mu \Vert \phi} H + \sfrac12
\Erw{\phi \Vert \phi}\pa^\mu H\big).
\eea
There is a common coupling constant $g$. 
The coupling parameters $c$, $c_5$, the structure constants
$f_{abc}$ and the overall coefficient $\mu$ in \eref{iniLVhiggs} are to be
determined by the resolution of obstructions at second order.

We shall see that the minimal interaction separately has a
non-resolvable obstruction at second order that can be cancelled by
the cross-terms with the self-interaction, giving ``chirality''
conditions on $c$ and $c_5$. We shall also see that the
self-interaction and the Higgs coupling both have non-resolvable
obstructions at second order that cancel each other when the Lie
algebra is $\mathfrak{su}(2)$ and the coefficient $\mu$ is fixed
appropriately. (Other Lie algebras should also be 
possible provided one admits several Higgs fields.) 

The cubic and quartic Higgs self-coupling will be fixed at third
order, as in \cite{AHM}. We shall also confirm that there are no
induced interactions at third order.

\subsection{Propagators and two-point obstructions.}

We shall use kinematic propagators throughout, with the only
exceptions being the local propagators of scaling dimension 4
\bea{propreno}
\eerw{T\pa_\mu H \pa'_\ka H'} &=& -\pa_\mu\pa_\ka T_{m_H} +
c_H\cdot \eta_{\mu\ka}\id, \notag \\
\eerw{TB_\mu B'_\ka} &=& -(\eta_{\mu\ka} + m^{-2}\pa_\mu\pa_\ka) T_m +
c_B \cdot m^{-2} \id,
\notag \\ 
\eerw{TF_{\mu\nu}F'_{\ka\la}} &=& - \pa_{[\mu}\eta_{\nu][\ka}\pa_{\la]} T_m
- c_F \cdot \eta_{\mu[\ka}\eta_{\la]\nu} \id.\eea
These propagators together with the respective Klein-Gordon equations,
the Dirac equation, and the equations of motion for the Proca fields
\eref{eomB} and \eref{eomA}
determine the relevant two-point obstructions, as listed in the
subsequent tables. All entries are
factors multiplying, or operators acting on $\id\equiv i \delta(x-x')$:
\bea{2pobsHD}
\begin{tabular}{l||c|c|}
Higgs & $H'$ & $\pa'^\ka H'$   \cr\hline\hline
$O_\mu(H,\cdot)$ & $0$ & $c_H\delta_\mu^\ka \noid $             \cr
$O_\mu(\pa^\mu H,\cdot)$ & $1$ & $-(1+c_H) \pa^\ka \noid$         \cr
\hline
  \end{tabular}
  \qquad\qquad
  \begin{tabular}{l||c|c|}
Dirac & $j'^\ka_b$ &$j'^{\ka5}_b$   \cr\hline\hline
$O_\mu(j^\mu_a,\cdot)$
    & $-f_{abc}j^\ka_c\noid$ & $-f_{abc}j^{\ka5}_c\noid$
            \cr
$O_\mu(j^{\mu5}_a,\cdot)$
    & $-f_{abc}j^{\ka5}_c\noid$ & $-f_{abc}j^\ka_c\noid$
            \cr
\hline
  \end{tabular}
  \eea
{\bf Table 4.} Two-point obstructions in the Higgs and Dirac sectors. (All
entries acting on $\id$)

\bea{2pobsPO20}
\hspace{-5mm}\begin{tabular}{l||c|c|c|c|}
Proca &$B'^{\ka}$&$\phi'$& $F'^{\ka\la}$ &$A'^\ka$   \cr\hline\hline
$O_{[\mu}(B_{\nu]},\cdot)$  &$-c_B m^{-2} \pa_{[\mu}\delta_{\nu]}^\ka$&$0$&$-c_F\delta_{\mu}^{[\ka}\delta^{\la]}_\nu$&$0$\cr
$O_\mu(B^{\mu},\cdot)$  &$-(1+c_B)m^{-2}\pa^\ka \noid$&$-m^{-2} \noid$&$0$&$0$\cr
$O_\mu(F^{\mu\nu},\cdot)$
&$-(1+c_B)\eta^{\nu\ka} \noid$&$-I'^\nu \noid$&  $-(1+c_F)\eta^{\nu[\ka}\pa^{\la]}\noid$&$-(\eta^{\nu\ka} - I'^\nu\pa^\ka)\noid$\cr
$O_{[\mu}(A_{\nu]},\cdot)$  &$0$&$0$&$-c_F\delta_{\mu}^{[\ka}\delta^{\la]}_\nu$&$0$\cr
$O_\mu(A^{\mu},\cdot)$  &$-I^\ka \noid$&$-(II')\noid$&$-I^{[\ka}\pa^{\la]}\noid$&$-(I^\ka-(II')\pa^\ka)\noid $\cr   $O_\mu(\phi,\cdot)$  &$-c_Bm^{-2}\delta_\mu^\ka$&$0$&$0$&$0$\cr  $O_\mu(w,\cdot)$  &$0$&$0$&$0$&$0$\cr
\hline
\end{tabular}
\eea
{\bf Table 5.} Two-point obstructions in the Proca sector. (All entries acting on $\id$)

Table 5 extends Table 1 for $F$ and $A$ in YM, despite the fact that
the fields are now massive.

\subsection{Adapting Lemmas from YM and QCD}

We adapt \lref{l:L1}--\lref{l:ww2}, allowing $L_1$ and $Q_1$ to depend
also on the field $\phi(c)$ as well as other string-independent fields
($B^\mu$ and $H$):

\begin{lemma}\label{l:L1-HK} If $\delta_cL_1(c)=\pa Q_1$ where
  $L_1$ is a   Wick polynomial in the fields $A_a, \phi_a$ and string-independent
  fields, and $Q_1$ is linear in $u_a$, then
\bea{L1} \hbox{\rm (i)}\quad \frac{\pa L_1}{\pa A_{\mu}} = \frac{\pa Q^\mu_1}{\pa w}, \qquad
\hbox{\rm (ii)}\quad \frac{\pa L_1}{\pa \phi} = \pa_\mu \Big(\frac{\pa
  Q^\mu_1}{\pa w}\Big), \qquad
\hbox{\rm (iii)}\quad \frac{\pa L_1}{\pa \phi} = \pa_\mu \Big(\frac{\pa L_1}{\pa A_{\mu}}\Big).
\eea
In particular, when $L_1$ does not contain $\phi_a$, then $\frac{\pa
  L_1}{\pa A_{a\mu}}$ is conserved.
\end{lemma}

{\em Proof:} The comparison of
$$\delta_cL_1=\Erw{\frac{\pa L_1}{\pa\phi}\Vert u}+\Erw{\frac{\pa L_1}{\pa A^\mu}\Vert\pa^\mu u}$$
with
$$\pa_\mu Q_1^\mu = \pa_\mu \Erw{\frac{\pa Q_1^\mu}{\pa u}\Vert u}
=\Erw{\pa_\mu \Big(\frac{\pa Q_1^\mu}{\pa u}\Big)\Vert u}+\Erw{\frac{\pa
  Q_1^\mu}{\pa u}\Vert \pa_\mu u} $$
immediately yields (i) and (ii). (iii) and the last statement are
obvious consequences of (i) and (ii).
\qed

As in QCD, define
\bea{w2} w_2(x,x'):=i[w',A'_\nu]\cdot I^\nu\delta_{xx'}.
\eea

\begin{lemma}\label{l:ww2-HK}  Both for YM and for QCD, $O^{(2)}\vert_{I\delta}$ is a total derivative. The
corresponding field $Q_2^\mu\vert_{I\delta}(x,x')$ arises from $Q^\mu_1(x)$ by a simple
replacement of $w(x)$ by $2w_2(x,x')$:
\bea{lem2} Q_2^\mu\vert_{I\delta}(x,x') = 2 \Erw{\frac{\pa Q^\mu_1}{\pa
    w}(x)\Vert w_2(x,x')}, \qquad\hbox{where}\quad w_2(x,x'):=i[w',A'_\nu ]\cdot I^\nu\delta_{xx'}.
\eea
\end{lemma}

The proof is almost identical with that of \lref{l:ww2}, just adding
the contribution from $O(F,\phi')$.

\begin{lemma}\label{l:aprio-HK}  \lref{l:aprio} (a priori cancellation
  of terms $\sim w_2$ at third order) holds unchanged also
  for HK.
\end{lemma}
(In the proof, there appear extra terms $\pm \Erw{w_2(x,x'')\Vert\frac{\pa L_2}{\pa\phi}}\cdot i\delta_{x,x'}$ in \eref{aprio1} and
\eref{aprio2}, which cancel each other.)

\lref{l:delta} and \lref{l:Opa} are general rules, not referring to a specific
model.

\subsection{$L$-$Q$ at second order}
\label{s:LQ2-HK}
At second order, we have to compute and resolve the obstructions:
\bea{O2-HK} O^{(2)}(x,x')= 2\Sy_{xx'} O_{Q_1}(L_1')\stackrel{!}=\delta_cL_2(x,c)\cdot \id-i\Sy_{xx'}\pa^x_\mu Q_2^\mu(x;x'),
\eea
where 
\lref{l:ww2-HK} already determines the part
$Q_2^\mu(x;x')\vert_{I\delta}$ resolving $O^{(2)}\vert_{I\delta}$. Thus, it remains to establish \eref{O2-HK} for the part $O^{(2)}\vert_\delta$ without
string-integrated delta functions.

\paragraph{LQ: Sector $J^2$.} Comes from min-min. Because $O_w(A)=0$,
\bea{LQ2-JJ} L_{2}(x)\vert_{J^2} &=& 0, \\ \notag
Q^\mu_{2}(x,x')\vert_{J^2} &=& 0.
\eea

\paragraph{LQ: Sector $P^2\cdot J$ (``Chirality'').} Comes from min-min, min-self and
self-min. By Table 4 and Table 5, min-self vanishes, and the  two contributions $\sim
\Erw{i[w,A]\Vert J}$ from the other two crossings cancel each other, provided 
the current $(c^2+c_5^2) j^\ka +
  2cc_5 j^{5\ka}$ from min-min equals the current $c j^\ka +
  c_5 j^{5\ka}$ from self-min. There are the non-chiral and two chiral solutions 
\bea{chi} (c=1,c_5=0), \quad \hbox{or}\quad (c=\sfrac12,c_5=\pm \sfrac 12),\eea
implementing the desired ``red'' cancellation. Then 
\bea{LQ2-PPJ} L_{2}(x)\vert_{P^2J} &=& 0, \\ \notag
Q^\mu_{2}(x,x')\vert_{P^2J\cdot\delta} &=&  0.
\eea

\paragraph{LQ: Sector $P\cdot J\cdot H$.} 
Comes from min-Higgs and Higgs-min. By Table 4 and Table 5, all
contributions vanish:
\bea{LQ2-PJH} L_{2}(x)\vert_{PJH} &=& 0, \\ \notag
Q^\mu_{2}(x,x')\vert_{PJH} &=& 0. 
\eea

\paragraph{LQ: Sector $P^3\cdot H$.} 
Comes from Higgs-self and self-Higgs. Table 5 yields exact cancellation: 
\bea{LQ2-PPPH} L_{2}(x)\vert_{P^3H} &=& 0, \\ \notag
Q^\mu_{2}(x,x')\vert_{P^3H\cdot\delta} &=& 0.
\eea

\paragraph{LQ: Sector $P^2\cdot H^2$.} 
Comes from Higgs-Higgs. Compare with [AHM]. 
\bea{LQ2-PPHH} L_{2}(x)\vert_{P^2H^2} &=& \sfrac{\mu^2}{m^2}\cdot
(1+c_B)\Erw{A_\mu\Vert A^\mu} H^2 + \mu^2\cdot\big(3a + \sfrac{m_H^2}{m^2}\big)\Erw{\phi\Vert\phi}H^2, \\ \notag
Q^\mu_{2}(x,x')\vert_{P^2H^2} &=& 2\sfrac{\mu^2}{m^2}\cdot
(1+c_B)\Erw{w\Vert A^\mu} H^2 \cdot\delta_{xx'}
\eea

\paragraph{LQ: Sector $P^4$ (``Lie algebra'').} 
Comes from Higgs-Higgs and self-self.
The contribution from Higgs-Higgs is
\bea{OHiggs} \notag
\mu^2\cdot
\big(-m_H^2\underbrace{\Erw{w\Vert\phi}\Erw{\phi\phi}}_{\delta_c\big(\frac14
\erw{\phi\vert \phi}\erw{\phi\vert \phi}\big)}
+ (1+c_H)\big(2\underbrace{(\Erw{\pa w\Vert\phi}\Erw{A\Vert\phi}+\Erw{A\Vert
  w}\Erw{A\Vert\phi})}_{\delta_c\big(\frac12 \erw{A\vert\phi}\erw{A\vert\phi}\big)}-\pa
[\Erw{w\Vert\phi}\Erw{A\vert\phi}]\big)\big)\cdot\id + O_{\rm
  Higgs},\eea
where 
$$O_{\rm Higgs}
=2\mu^2\cdot(\Erw{w\Vert\phi}\Erw{B\Vert A}-\Erw{w\Vert
  B}\Erw{\phi\Vert A}) \cdot\id$$
is a non-resolvable obstruction%
\footnote{This is a substantial correction vs archive version! It
  yields no condition on $c_H$.
    It holds $\erw{w\vert\phi}\erw{B\vert A}-\erw{w\vert
  B}\erw{\phi\vert A} = \frac12\delta_c(\erw{\phi\vert\phi}\erw{B\vert
  A}-2\erw{\phi\vert B}^2) -\frac12 \pa (\erw{\phi\vert\phi}\erw{B\vert w})$, but $\erw{\phi\vert\phi}\erw{B\vert A}$ and $\erw{\phi\vert B}^2$ of dimension 5 and 6 are not
admissible as induced interactions.}. $O_{\rm Higgs}$ must be cancelled by
the contribution from self-self.

The contribution from self-self is%
\footnote{Substantial correction vs archive version! No condition on $c_B$.}
$$(1+c_F)\Big[-\sfrac12\delta_c\big(\Erw{i[A_\ka,A_\la]\Vert
    i[A^\ka,A^\la]}\big)\cdot \id+2\Sy_{xx'}\pa_\ka\big(\Erw{ i[w,A_\la]\Vert i[A^\ka,A^\la]}\cdot\id\big)\Big]
$$
$$ +(1+c_B)m^2 \Big[\sfrac14\delta_c\big(\Erw{i[\phi,A]\Vert i[\phi,A]}\big) - \Sy_{xx'}\pa_\ka\big(\sfrac12\Erw{i[\phi,w]\Vert
i[\phi,A^\ka]}\cdot\id\big)\Big] + O_{\rm
self} ,$$
with the non-resolvable obstruction
$$O_{\rm self} = -\sfrac12 m^2\Erw{i[w,A_\ka]\Vert i[\phi,B^\ka]}
\cdot\id.$$

We observe:

(i) $c_H$ and $c_B$ are  not fixed at second order. At third order,
they must be put $=-1$ because the terms $\sim \erw
{A\vert \phi}^2$ and $\sim \erw{i[A,\phi]\vert i[A,\phi]}$ in $L_2$
and $\sim \erw{w\vert\phi}\erw{A\vert\phi}$ and $\sim \erw{i[w,\phi]\vert
  i[A,\phi]}$ in $Q_2\vert_\delta$ would produce non-resolvable obstructions. \\
(ii) $O_{\rm self}$
cancels $O_{\rm Higgs}$ (the desired ``green''
cancellation):
$$ 2\mu^2\cdot (\Erw{w\Vert\phi}\Erw{B\Vert A}-\Erw{w\Vert
  B}\Erw{\phi\Vert A}) - \sfrac12 m^2\Erw{i[w,A_\ka]\Vert i[\phi,B^\ka]} \stackrel!=0,$$
provided the structure constants of the Lie algebra are  $\pm2\mu
m\inv\cdot\eps_{abc}$, hence the Lie algebra must be $su(2)$ (for the
field content at hand).
With the convention for the generators $\tau_a=\frac12\sigma_a$:
\bea{fepsmu}
f_{abc}=\eps_{abc}, \qquad \mu=\sfrac12m,
\eea
where the sign of $\mu$ can be adjusted by redefining $H\to -H$.


We conclude
\bea{LQ2-PPPP} L_{2}(x)\vert_{P^4} &=& \mu^2\cdot \big(-\sfrac14m_H^2\Erw{\phi\Vert\phi}\Erw{\phi\Vert\phi}\big)-\sfrac12(1+c_F)
\Erw{i[A_\ka,A_\la]\Vert i[A^\ka,A^\la]} \\ \notag
&&+ \sfrac14(1+c_B)m^2 \Erw{i[A,\phi]\Vert i[A,\phi]}
+\sfrac12(1+c_H)\Erw{A\vert\phi}^2,
\\ \notag
Q^\mu_{2}(x,x')\vert_{P^4\cdot\delta} &=& 
- 2(1+c_F)\Erw{i[A^\mu,A^\nu]\Vert i[w,A_\nu]}\cdot\delta_{xx'} \\ \notag
&&+ \sfrac12(1+c_B)m^2 \Erw{i[w,\phi]\Vert i[A,\phi]}\cdot\delta_{xx'} +(1+c_H)\Erw{w\vert\phi}\Erw{A\vert\phi}\cdot\delta_{xx'},
\eea

\paragraph{LQ: Sector $H^4$.} As in the abelian model, we are free to add
the local and renormalizable term 
\bea{LQ2-HHHH} L_2(x)\vert_{H^4} &=& \mu^2\cdot
m^{-2}\cdot bH^4, \\ \notag Q_2^\mu(x,x')\vert_{H^4} &=&0.
\eea

\paragraph{LQ: Total at second order.}
Collecting all terms, we have  
\bea{L2-HK} L_2(x)&=&\underbrace{-\sfrac12(1+c_F)
\Erw{i[A_\ka,A_\la]\Vert i[A^\ka,A^\la]}}_{=:L_{2,\rm  self}}+\underbrace{\sfrac{m^2}4
  \big(\big(3a +
  \sfrac{m_H^2}{m^2}\big)\Erw{\phi\Vert\phi}H^2-\sfrac{m_H^2}4\Erw{\phi\Vert\phi}\Erw{\phi\Vert\phi}\big)
+ \sfrac14bH^4}_{=:L_{2,\rm
  Higgs}}
\notag \\
&&+\underbrace{(1+c_H)\cdot\sfrac{m^2}4\Erw{A\Vert\phi}\Erw{A\Vert\phi}+(1+c_B)\cdot\big(\sfrac{m^2}4\Erw{i[A,\phi]\Vert
    i[A,\phi]} + \sfrac{\mu^2}{m^2}\Erw{A_\mu\Vert A^\mu}H^2\big)}_{=:L_2^*},
\eea
and 
\bea{Q2-HK}Q^\mu_{2}(x,x') = Q^\mu_{2}\vert_{I\delta} + Q^\mu_{2,\rm
self} +Q^{*\mu}_{2}  = 
2 \erw{\frac{\pa Q^\mu_1}{\pa w}\vert w_2(x,x')} + \erw{\frac{\pa
    L_{2,\rm self}}{\pa A_\mu}\vert w} \cdot\delta_{xx'}+\erw{\frac{\pa
    L^*_2}{\pa A_\mu}\vert w} \cdot\delta_{xx'}.\eea
The string delta part is as specified by
\lref{l:ww2-HK}, and the delta part again (as in QCD) satisfies \eref{Q2L2}.

\subsection{$L$-$Q$ at third order}

At third order, we have to compute and resolve the obstructions
\eref{O3}:
\bea{O3-HK} \notag O^{(3)} = 3\Sy_{xx'x''}\big(O_{Q_2}(L_1'') + O_{Q_1}(L_2')\cdot
\delta_{x'x''}\big)\stackrel!= L_3\cdot i\delta_{xx'x''} -
i\Sy_{xx'x''}\pa_\mu Q^3(x;x',x'').
\eea
The computations are analogous to QCD, \sref{s:QCD}. First, 
\lref{l:aprio-HK} reduces the third-order obstruction to
\bea{O3remain-HK}
O^{(3)}\stackrel{\rm der}=3\Sy_{xx'x''}\big(\Erw{\frac{\pa Q_2}{\pa
    w_2}\Vert O(w_2,L_1'')} + O_{Q_2\vert_\delta}(L_1'')\vert_{I\delta}+ O_{Q_2\vert_\delta}(L_1'')\vert_\delta+ O_{Q_1}(L_2')\vert_\delta\cdot\delta_{x'x''} \big).
\eea
The first two terms are treated as in \sref{s:QCD}. With Option (A),
they are separately derivatives with a sharp string, provided the
sofar missing propagators are renormalized as in \eref{TpaAX}
with \eref{OAX} for $X'=A'$, $F'$ and
\bea{OAX-HK}
O_\mu(A_\nu,B'^\ka) &:=& -\sfrac14 \eta_{\mu\nu} I^\ka\id 
\\ \notag O_\mu(A_\nu,\phi') &:=& -\sfrac14 \eta_{\mu\nu}
(II')\id.\eea
It is again checked that these definitions are consistent with the
respective lines of Table 5, which were computed by using the equations of motion
\eref{eomB} and \eref{eomA} under the $T$-product.

Namely, $w_2=0$ and similar arguments as in the proof of \pref{p:axial} apply also for
$X'=B'$ and $X'=\phi'$. Specifically, the contribution from $X'=\phi'$
makes up for the fact that $\frac{\pa L'}{\pa A'^\ka}$ in the
contribution from $X'=A'$ is no
longer conserved, by \lref{l:L1-HK}. For smeared strings, \cjref{cj:optA} applies
again.

With Option (B), the two terms also vanish because $w_2$ and $Q_2\vert_\delta$ are ``inert''.

Therefore, the resolution of $O^{(3)}$ reduces to that of the
unproblematic last two terms (without string deltas) in \eref{O3remain-HK}, which can be computed
with Table 4 and 5, supplemented by \eref{OAX} and \eref{OAX-HK}. 


Here, it is convenient%
\footnote{$c_B\neq-1$ and $c_H\neq-1$ separately produce non-resolvable obstructions at third
  order, unless $c_B=c_H$. On the other hand, $c_F$ seems to remain arbitrary, as $c$ in
  scalar QED. The choice $c_F=0$ corresponds to $L_2$ in gauge
  theory.}
to put
\bea{cHB}c_H=-1, \qquad c_B=-1.
\eea
Then, the only obstructions have  field content
$u\phi^3H$ and $u\phi H^3$. Their absence fixes the ``shape of the Higgs potential'', as in the abelian model:
\begin{prop}\label{p:HKab}  With \eref{cHB}, the term
  $O_{Q_2\vert_\delta}(L_1'')\vert_\delta$ in \eref{O3remain-HK}
  vanishes identically, and $O_{Q_1}(L_2')\vert_\delta$ vanishes if
  and only iff the parameters of 
the Higgs self-coupling $aH^3+bH^4$ in \eref{iniLVhiggs} and \eref{LQ2-HHHH} are
$$a=-\frac12 \frac{m_H^2}{m^2}, \qquad b=-\frac14
\frac{m_H^2}{m^2}.$$
With these values the Higgs potential becomes the shifted double-well
potential; and  $L_3=Q_3^\mu=0$.
\end{prop}

{\em Proof:} With \eref{cHB}, $Q_2\vert_\delta$ is a Wick polynomial in $w$ and
$A$ only, where $w$ is inert and $O(A,X')$ has no contributions without
string-deltas, except $O(A,F')$ (by scaling dimension). This contribution is computed as in
the proof of \pref{p:L3-QCD}, and vanishes thanks to the Jacobi
identity. Therefore, $O_{Q_2\vert_\delta}(L_1'')\vert_\delta=0$.

$O_{Q_1}(L'_{2,\rm self})\vert_\delta$ arises only from $O(F,A')$,
which produces a multiple of
$$\Erw{i[w,A_\nu]\Vert i[A^\la,i[A_\ka,A_\la]]}\cdot \eta^{\nu\ka}\id
\stackrel{\rm Jacobi}= \sfrac12 \Erw{i[w,i[A^\ka,A^\la]] \Vert i[A_\ka,A_\la]}\cdot\id=0$$
by \eref{cart-inv}. By Table 5, $O_{Q_{1,\rm min}}(L'_{2,\rm Higgs})=0$, and
$O_{Q_{1,\rm self}}(L'_{2,\rm Higgs})$ is a multiple of $\erw{i[w,\phi]\vert\phi}\cdot\id=0$.

It remains to study $O_{Q_{1,\rm Higgs}}(L'_{2,\rm Higgs})$, where
$Q_{1,\rm Higgs}$ and $L'_{2,\rm Higgs}$ is the same expressions as in the
abelian model \cite{AHM}.  Therefore, its vanishing fixes the
values of $a$ and $b$, exactly as in the abelian model \cite{AHM}. \qed


Let us now look at additional contributions that arise when
$1+c_B\neq 0$ or $1+c_H\neq 0$. In $$O_{Q_1}(L_{2,\rm self +
  Higgs}(A',\phi',H'))\vert_\delta + O_{Q_{2,\rm self}(w,A)}(L_1'')\vert_\delta,$$ there are no
such terms because all pertinent two-point obstructions are independent of $c_B$ and $c_H$.

Thus, we have to study
$$O_{Q_1}(L_{2}^*(A',\phi',H'))\vert_\delta + O_{Q_2^*(w,A,\phi,H)}(L_1'')\vert_\delta.$$
Because $L_2'^*=L_2^*(A',\phi',H')$ and
$Q_2^* = Q_2^*(w,A,\phi,H)$, only $O(F,A')$, $O(B,\phi')$, $O(\pa H,H')$ and
$O(\phi,B')$, $O(H,\pa'H')$ contribute to the first, resp.\ second term.

Here, it is convenient to use \eref{fepsmu} and write: 
\bea{L2*}L_2^* = (1+c_B)\erw{A\vert A}\big(\sfrac{m^2}4\erw{\phi\vert\phi}+\sfrac14H^2\big)
+(c_H-c_B)\sfrac{m^2}4\erw{A\vert\phi}\erw{A\vert\phi},
\eea
and similar for $Q_2^*\vert_\delta$. 
We find 
\bea{} \notag O_{Q_1}(L_{2}^*(A',\phi'))\vert_\delta\delta_{x'x''} &\!\!\!\!=\!\!\!\!& -
   (c_H-c_B)\sfrac{m^2}4 \Erw{i[w,A]\Vert \phi}\Erw{A\Vert \phi} \, i\delta_{\rm tot}- 
  (c_H-c_B)\sfrac{m}4\Erw{w\Vert A}\Erw{A\Vert \phi}H\,
  i\delta_{\rm tot},\\\notag
  O_{Q_2^*(w,A,\phi)}(L_1'')\vert_\delta &\!\!\!\!=\!\!\!\!& -c_B(1+c_B)
\sfrac{m}4\Erw{w\Vert A}\Erw{\phi\Vert A}H \, i\delta_{\rm tot} +
c_H(1+c_B)\sfrac m4\Erw{w\Vert A}\Erw{A\vert \phi}H \,i\delta_{\rm tot} \\ \notag &\!\!\!\!\!\!\!\!& -c_B
(c_H-c_B)\big(\sfrac{m^2}8\Erw{w\Vert i[A,\phi]}\Erw{A\Vert\phi} +
\sfrac m8\Erw{w\Vert A}\Erw{A\Vert\phi}H +\sfrac m8\Erw{A\Vert
  A}\Erw{w\Vert\phi}H\big) i\delta_{\rm tot}.
\eea
The total is
$$-\sfrac m8\big[(2+c_B)(c_H-c_B)m \Erw{w\Vert i[A,\phi]}\Erw{A\Vert\phi}
+c_B(c_H-c_B) \big(\Erw{w\Vert\phi}\Erw{A\Vert A}-\Erw{w\Vert A}\Erw{A\Vert\phi}\big)H\big]i\delta_{\rm tot}$$
with three linearly independent non-resolvable
obstructions. Putting their coefficients to zero, one finds the condition
$$c_B=c_H.$$
(In the case of the abelian Higgs model, there is no condition at all, see \cite[Footnote~3]{AHM}).

\paragraph{LQ: Higher orders $n>3$.} Higher-order interactions $L_n$
are determined by the parts of $O^{(n)}$ without string-integrated
delta functions. At fourth order,
$O_{Q_1}(L_3)\vert_\delta=0$ because $L_3=0$; $O_{Q_3}(L_1''')\vert_\delta=0$ because
$Q_3\vert_\delta=0$; and $O_{Q_2\vert_\delta}(L_2'')=0$ because
$Q_2\vert_\delta$ and $L_2$ are Wick polynomials in ($w$ and) $A$, and
$O(A,A')\vert_\delta=0$ by \eref{OAX}. This can presumably be
generalized to all orders by induction, where the hardest task is to
establish resolvability of the parts of $O^{(n)}$ with string-integrated
delta functions, which must be total derivatives, i.e., the existence of $Q_n\vert_{I\delta}$.

\subsection{$L$-$V$ at second order}
At second order, we have to compute and resolve the obstructions \eref{P2}.

We found the same constraints on propagators ($c_B=c_H=-1$) and on
parameters (chirality and Lie algebra) as in $L$-$Q$. The induced
interaction $L_2$ is the same as in $L$-$Q$. In addition, we found
\bea{KU2-HK}
K_{2,\rm QCD} &=& -\sfrac12(1+c_F) \Erw{i[B_\mu,B_\nu]\Vert i[B^\mu,B^\nu]}
-\sfrac2m \Erw{F^{\mu\nu}\Vert
  i[B_\mu,B_\nu]}H -\sfrac34\Erw{B_\mu\Vert
  B^\mu}H^2-\sfrac{m_H^2}{16m^2}H^4 \quad
\notag\\ &&-\sfrac2m \Erw{B_\mu\Vert J^\mu}H-\sfrac1{m^2}\Erw{J_\mu\Vert J^\mu},
\eea
and a very lengthy expression for $U_2^\mu$ with contributions in all
sectors except $J^2$ and $H^4$.

\subsection{Towards the full weak interactions}
In an ongoing work \cite{weak}, we are dealing with the full weak
interaction (massive vector bosons with mass ratio
$M_W/M_Z=\cos\theta_W$ given by the
Weinberg angle, one generation of leptons with different masses, and
the photon and the Higgs).

The analysis in the fermionic sector (containing vector and axial
currents as well as scalar and pseudoscalar fields $S_a:=\ol\psi \tau_a \psi$
and $S^5_a:=\ol\psi \tau_a \gamma^5\psi$) was completed in
\cite{GMV}, with the main result being the prediction of chirality
from string-independence. It remains to consider the bosonic sector.

In \cite{weak}, we are predicting the bosonic sector
of a class of models including the electroweak theory. The input is any number of massive
particles of spin 1 (``MVBs'') and massless particles of helicity $\pm 1$
(``photons'') and one massive scalar 
particle (``Higgs''). The masses are assumed to be given. String-independence is
postulated at first, second and third order.

At first order, SI (via the LQ condition)
\\ (i) classifies the structure of cubic
self-interactions of massive and massless vector bosons,\\
(ii) constrains the masses $m_a$ of MVBs coupled to photons, and
\\
(ii) classifies the structure of cubic interactions between MWBs and the Higgs.

At second and third order, SI (via the necessary resolution of obstructions)
\\ (i) constrains the self-coupling coefficients, which must be the
structure constants of a reductive Lie algebra, \\
(ii) relates the coupling coefficients with the Higgs to the masses,
\\
(iii) relates the structure constants $f_{abc}$ of the Lie algebra to the
masses and the coupling coefficients $k_a$ with the Higgs, and
\\
(iv) determines the ``induced'' quartic interactions.

Specifically for the $W$ and $Z$ bosons and the photon with their
given masses of particle physics, it follows from SI that $m_W\leq
m_Z$, which allows to introduce an  angle
$\Theta$ (the Weinberg angle): $\cos\Theta := m_W/m_Z$. Furthermore,
$$k_W/k_Z = m_W^2/m_Z^2 = \cos^2\Theta, $$
$$(f_{12A})^2/(f_{12Z})^2 = \tan^2\Theta.$$
These are precisely the relations known from the textbooks, where the
gauge group $U(1)\times SU(2)$ with gauge coupling constants $g_1$ and
$g_2$ is said to be spontaneously broken, such that $\frac12\tau_0+\tau_3 =
\frac12(\eins+\sigma_3)$ is the generator of the unbroken subgroup,
with the Weinberg angle defined by
$$\tan\Theta = \sfrac{g_1}{g_2}.$$

The conventions in \sref{s:HK} are different from those in
\cite{GMV} which will also be used in \cite{weak}. 
The coupling constant in
\cite{GMV} is minus the present coupling constant as in \sref{s:HK},
which can be identified with the $SU(2)$ coupling constant $g_2$. In
gauge theory, the minimal couplings are
\bea{gaugemin}
L_{\rm min} = g_1 W^0_\mu(\ol \ell \gamma^\mu\pi(\tau_0)\ell) + g_2\sumno_{i=1,2,3}W^i
_\mu(\ol \ell \gamma^\mu\pi(\tau_i)\ell),
\eea
where $\ell = \bpm \nu \\[-2mm] e\epm$ is the lepton doublet, $\tau_0=\eins$ and $\pi_{L/R}(\tau_0)=y_{L/R}$ are the
hypercharges ($y_R(\nu)=0$, $y_R(e)=-2$, $y_L(\ell)=-1$), and
$\pi_L(\tau_i)=\frac12\sigma^i$ and $\pi_R(\tau_i)=0$.
Putting $g_1= \frac12 g_2\tan\theta_W$ and 
$$W^\pm = \sfrac1{\sqrt 2}(W^1\mp W^2), \qquad A=
W^0\cos\theta_W+W^3\sin\theta_W, \qquad Z=
-W^0\sin\theta_W+W^3\cos\theta_W,$$
one arrives at the minimal interactions with electrons and neutrinos
\bea{Lminweak}
L_{\rm min}^{\rm weak} &=& g\Big(\sin\theta_W\cdot A_\mu (\ol e\gamma^\mu e)+
\frac1{2\cos\theta_W}\cdot Z_\mu\big[(1-2\sin^2\theta_W)(\ol e_L\gamma^\mu e_L) -2\sin^2\theta_W (\ol e_R\gamma^\mu e_R)- (\ol \nu_L\gamma^\mu \nu_L)\big]\notag \\ && - \frac1{\sqrt2}\cdot\big[W^-_\mu(\ol
e_L\gamma^\mu\nu_L)+W^+_\mu(\ol \nu_L\gamma^\mu e_L)\big]\Big)\eea
in accord with
\cite[Thm.~8]{GMV}, with the identifications
\bea{ggg}
g_2=-g, \qquad g_1= -g\tan\theta_W.
\eea
The unit of electric charge is
$e=g_1g_2/\sqrt{g_1^2+g_2^2} = g_2\tan\Theta$.

\newpage

In the Higgs-Kibble model for $SU(2)$ with coupling constant $g_2$ and
$\pi(\tau_i)=\tau_i$, there is no hypercharge, and the $g_2$-term of \eref{gaugemin} goes
along with $g_2L_1^{\rm self}=-\frac12 g_2
\sum_{abc=1,2,3}\eps_{abc} F^aW^bW^c+\dots$, which explains the sign
discrepancy
$g_2=-g$ in \cite{GMV}. Finally, with $W^3 = A\sin\theta_W + Z\cos\theta_W$, one
gets
$$S_1^{\rm self}\equiv gL_1^{\rm self}= g\sum_{abc}f_{abc}  F^aW^bW^c
+ \dots,$$
where now $a,b,c$ run over the values $1,2,3=Z$ and $4=A$ and
$$f_{123} = \sfrac12 \cos\theta_W, \quad f_{124} = \sfrac12
\sin\theta_W,$$
in accord with the convention in \cite[Sect.~4]{GMV}. The masses and
Higgs coupings are exactly the same as the ones predicted by
string-independence in \cite{weak}.

\medskip

For more general models, \cite[Sect.~7]{weak} expresses the conditions of string
independence (in the bosonic sector) as a system of matrix
equations (on the space of massive vector bosons), that may be
interpreted as a condition on the Higgs couplings, arising by a deformation of the adjoint
representation of the ``massive'' generators of the underlying Lie
algebra in terms of the masses of the massive vector bosons.


\begin{thebibliography}{9} \itemsep-1mm
\bibitem{Cha23} M. Chantreau: Master Internship report, Lyon--Göttingen 2023
 \bibitem{GassPhD} C. Gass: PhD thesis, Göttingen 2022
 \bibitem{YM} C. Gass, J.M. Gracia-Bond\'ia, J. Mund: Revisiting
    the Okubo–Marshak Argument, Symmetry 13 (2021) 1645.
  \bibitem{GMV} 
J.M. Gracia-Bond\'ia, J. Mund, J.C. V\'arilly: The chirality
theorem. Ann. Henri Poincaré 19 (2018) 843--874.
\bibitem{weak} J.M. Gracia-Bond\'ia,  K.-H. Rehren, J.C. V\'arilly: The full electroweak interaction:
an autonomous account, arXiv:2409.10668.
\bibitem{Hemp23} I. Hemprich: Master's lab course report, Göttingen 2023.
 \bibitem{Gauss} J. Mund, K.-H. Rehren, B. Schroer:  Gauss’ Law and
   string-localized quantum field theory. JHEP 01 (2020) 001.
 \bibitem{Infra} J. Mund, K.-H. Rehren, B. Schroer:  Infraparticle
      quantum fields and the formation of photon clouds. JHEP 04
      (2022) 083. 
\bibitem{AHM} J. Mund, K.-H. Rehren, B. Schroer: How the Higgs
        potential got its shape. Nucl.\ Phys.\ B987 (2023) 116109.
\bibitem{Pab24} M. Pabst: Bachelor's thesis, Göttingen 2024
\bibitem{LV} K.-H. Rehren: On the effect of derivative interactions in
  quantum field theory. arXiv:2405.09168.
  \bibitem{aut} K.-H. Rehren, L.T. Cardoso, C. Gass, J.M. Gracia-Bond\'ia,
    B. Schroer, J.C. V\'arilly: sQFT: an autonomous explanation of
    the interactions of quantum particles. Found. Phys. 54 (2024) 57, arXiv:2405.09366. 
\bibitem{Tipp} F. Tippner: Bachelor's thesis 2019, Göttingen
  
 \end{thebibliography}
\end{document}